\documentclass[twocolumn, astrosymb]{aastex631}
\usepackage{graphicx}
\usepackage{amsmath}
\usepackage{amssymb}
\usepackage{dirtytalk}
\usepackage{threeparttable}
\usepackage[figuresright]{rotating}
\shorttitle{The nebular properties of star-forming galaxies from LEGA-C}
\shortauthors{Helton et al.}

\begin{document}

%% Title of the paper.
\title{The nebular properties of star-forming galaxies at \\ intermediate redshift from the Large Early Galaxy Astrophysics Census}

%% Authors of the paper
\author[0000-0003-4337-6211]{Jakob M. Helton}
\affiliation{Steward Observatory, University of Arizona, 933 N. Cherry Ave., Tucson, AZ 85721, USA}
\affiliation{Department of Astrophysical Sciences, Princeton University, 4 Ivy Lane, Princeton, NJ 08544, USA}

\author[0000-0001-6369-1636]{Allison L. Strom}
\affiliation{Department of Astrophysical Sciences, Princeton University, 4 Ivy Lane, Princeton, NJ 08544, USA}
\affiliation{Department of Physics and Astronomy and Center for Interdisciplinary Exploration and Research in Astrophysics (CIERA), Northwestern University, 2145 Sheridan Road, Evanston, IL 60208, USA}

\author[0000-0002-5612-3427]{Jenny E. Greene}
\affiliation{Department of Astrophysical Sciences, Princeton University, 4 Ivy Lane, Princeton, NJ 08544, USA}

\author[0000-0001-5063-8254]{Rachel Bezanson}
\affiliation{Department of Physics and Astronomy, University of Pittsburgh, 3941 O'Hara St., Pittsburgh, PA, 15260, USA}

\author[0000-0002-1691-8217]{Rachael Beaton}
\affiliation{Department of Astrophysical Sciences, Princeton University, 4 Ivy Lane, Princeton, NJ 08544, USA}
\affiliation{The Observatories of the Carnegie Institution for Science, 813 Santa Barbara St., Pasadena, CA 91101, USA} 

%% Abstract of the paper.
\begin{abstract}
We present a detailed study of the partial rest-optical ($\lambda_{\mathrm{obs}} \approx 3600-5600\,$\AA) spectra of $N = 767$ star-forming galaxies at $0.6 < z < 1.0$ from the Large Early Galaxy Astrophysics Census (LEGA-C). We compare this sample with low-redshift ($z \sim 0$) galaxies from the Sloan Digital Sky Survey (SDSS), intermediate-redshift ($z \sim 1.6$) galaxies from the Fiber Multi-Object Spectrograph (FMOS)-COSMOS Survey, and high-redshift ($z \sim 2$) galaxies from the Keck Baryonic Structure Survey (KBSS). At a lookback time of $6-8\ \mathrm{Gyr}$, galaxies with stellar masses $\mathrm{log}(\mathrm{M_{\ast}/M_{\odot}}) > 10.50$ appear remarkably similar to $z \sim 0$ galaxies in terms of their nebular excitation, as measured using [O\,III]$\lambda5008$/H$\beta$. There is some evidence that $0.6 < z < 1.0$ galaxies with lower $\mathrm{M_{\ast}}$ have higher [O\,III]$\lambda5008$/H$\beta$ than $z \sim 0$ galaxies and are more similar to less evolved $z \sim 1.6$ and $z \sim 2$ galaxies, which are offset from the $z \sim 0$ locus at all $\mathrm{M_{\ast}}$. We explore the impact of selection effects, contributions from active galactic nuclei, and variations in physical conditions (ionization parameter and gas-phase oxygen abundance) on the apparent distribution of [O\,III]$\lambda5008$/H$\beta$ and find somewhat higher ionization in $0.6 < z < 1.0$ galaxies with lower $\mathrm{M_{\ast}}$ relative to $z \sim 0$ galaxies. We use new near-infrared spectroscopic observations of a subsample of LEGA-C galaxies to investigate other probes of enrichment and excitation. Our analysis demonstrates the importance of obtaining complete rest-optical spectra of galaxies in order to disentangle these effects.
\end{abstract}

%% Defines the keywords.

\let\svthefootnote\thefootnote
\newcommand\freefootnote[1]{
  \let\thefootnote\relax
  \footnotetext{#1}
  \let\thefootnote\svthefootnote}
\freefootnote{Corresponding author: Jakob M. Helton \\ \href{mailto:jakobhelton@email.arizona.edu}{jakobhelton@email.arizona.edu}}

%%%%%%%%%%%%%%%%%%%%%%%%%%%%%%%%%%%%%%%%%%%%%%

%%%%%%%%%%%%%%% BODY OF PAPER %%%%%%%%%%%%%%%%

%%%%%
\section{Introduction}
\label{SectionOne}

Surveys of local galaxies ($z \sim 0$) show that there is remarkable diversity in the properties of the galaxy population, including a broad range of masses, star formation rates (SFRs), morphologies, kinematics, and chemical composition \citep[for a review, see][]{Blanton:2009}. Distant galaxies ($z \sim 2$) also exhibit remarkable diversity, but have a different range of observed properties, with galaxies at fixed stellar mass ($\mathrm{M}_{\ast}$) having higher SFRs and smaller physical sizes than their present-day counterparts \citep[e.g.,][]{Whitaker:2012, Law:2012, Tacconi:2013}. These differences with redshift correspond to a notable change in cosmic star formation rate density, which peaked around $z \sim 2$ and has been declining ever since \citep{Madau:1998, Madau:2014}.  \par

The complex astrophysics that cause galaxies to transition from highly star-forming to having declining or ``quenched" star formation histories can be investigated using the spectra of galaxies observed during the $10-11$~Gyr between $z \sim 2$ and $z \sim 0$. The strong emission lines in the rest-optical ($\lambda_{\mathrm{rest}} \approx 3000-7000$\AA) spectra of galaxies are sensitive to the origin and physical state of the ionized gas and can reveal information about the massive stars and gas in H\,II regions, as well as the presence of active galactic nuclei or shocked gas. Diagnostics based on the relative strength of these emission lines are commonly used to determine the density \citep[][]{Dopita:1976, Keenan:1992, Kewley:2019a}, enrichment \citep[][]{Pagel:1979, Baldry:2002, Tremonti:2004}, excitation \citep[][]{Baldwin:1981, Kewley:2013a, Kewley:2013b}, and ionization state \citep[][]{Penston:1990, Dopita:2000, Moustakas:2010} of gas in galaxies at all redshifts \citep[for a review, see][]{Kewley:2019}.  \par

Low-redshift star-forming galaxies ($z \sim 0$; present-day) form a relatively narrow sequence in common emission line diagrams that compare different line ratios with one another or with global properties of galaxies such as $\mathrm{M}_{\ast}$ \citep[e.g.,][]{Kauffmann:2003, Juneau:2011}. These sequences arise due to strong correlations between the properties of galaxies that are responsible for their nebular spectra, especially gas-phase metallicity and ionization. High-redshift galaxies ($z \sim 2$; corresponding to a lookback time of $\sim 10-11$~Gyr) form similar sequences in these diagrams \citep[e.g.,][]{Steidel:2014, Shapley:2015, Strom:2017, Runco:2021}; however, they occupy distinct regions of parameter space when compared to low-redshift galaxies. This observed difference indicates that the nebular conditions in galaxies evolve significantly with redshift \citep[][]{Kewley:2013b, Kewley:2013a}. Differences in both the properties of stellar populations (e.g., metallicity, shape of the ionizing radiation field, and ionizing photon output) and the properties of the interstellar medium (ISM; e.g., gas-phase metallicity, electron temperature, electron density) seem to be responsible for this evolution \citep{Liu:2008, Masters:2014, Steidel:2016, Topping:2020, Sanders:2021, Strom:2022}. \par

Studies of intermediate-$z$ galaxies offer a more complicated picture. At $z \sim 1.4 - 1.7$ (corresponding to a lookback time of $\sim 9-10$~Gyr), galaxies have higher metallicity and lower excitation on average than $z\sim2$ galaxies at fixed $\mathrm{M}_{\ast}$ \citep[e.g.,][]{Zahid:2014, Kashino:2019, Topping:2021}. At the same time, there are still significant differences with respect to the local population, particularly at low $\mathrm{M}_{\ast}$. At $0.3 < z <1.0$ (corresponding to a lookback time of $\sim 3-8$~Gyr), most studies still find lower metallicities at fixed $\mathrm{M}_{\ast}$ than $z \sim 0$ galaxies \citep{Lamareille:2009, Juneau:2011, Zahid:2011}. However, there is disagreement in how these changes manifest as a function of $\mathrm{M}_{\ast}$ (with higher mass galaxies potentially evolving more rapidly) and whether changes in enrichment are solely responsible for the differences in galaxies' spectra. \par

Historically, intermediate-redshift galaxies have been more difficult to place in context with the much larger samples of $z \sim 0$ and $z \sim 2$ galaxies with complete rest-optical spectra that enable the effects of, e.g., enrichment and ionization to be disentangled. Using optical ($\lambda_{\mathrm{obs}} \approx 3600-7000\,$\AA) spectrographs alone, surveys are limited to $z \lesssim 0.5$; at higher redshifts, important lines tracing SFR and enrichment (including [N\,II]$\lambda6585$ and H$\alpha$) shift into the near-infrared (NIR; $\lambda_{\mathrm{obs}} \approx 7000-24000\,$\AA). Conversely, surveys using NIR spectrographs are limited to $z \gtrsim 1.6$; at lower redshifts, other features, including the density-sensitive [O\,II]$\lambda\lambda3727,3729$ doublet, are inaccessible at NIR wavelengths. Thus, complete rest-optical spectroscopic studies of intermediate-$z$ galaxies currently require substantial investment in both optical and NIR observations. \par

In coming years, it will be possible to obtain complete rest-optical spectra for large samples of intermediate-redshift galaxies more efficiently, using instruments such as the Prime Focus Spectrograph on the Subaru Telescope \citep[PFS,][]{Takada:2014} and the Multi-Object Optical and Near-infrared Spectrograph on the European Southern Observatory (ESO) Very Large Telescope \citep[VLT, MOONS,][]{Taylor:2018}, which are designed to simultaneously observe in the optical and NIR. However, we do not need to wait for future instrumentation to begin making progress toward the critical goal of understanding how the spectroscopic properties of galaxies have evolved over the last $10-11$~Gyr. This paper presents the first results from the PFS Pathfinder (Strom et~al., in prep.), an ongoing observational program designed to assemble complete rest-optical spectra of intermediate-$z$ galaxies using separate optical and NIR observations of the same galaxies. These data will ultimately be used to investigate the full rest-optical nebular properties of galaxies spanning $0.5 < z < 1.6$. \par

Here, we consider the nebular properties of a sample of $0.6 < z < 1.0$ galaxies drawn from the Large Early Galaxy Astrophysics Census \citep[LEGA-C,][]{vanDerWel:2016}, which have existing high-quality optical spectra from VLT/VIMOS and new NIR spectra from Keck/MOSFIRE and Magellan/FIRE. This paper proceeds as follows. In Section~\ref{SectionTwo}, we describe the various observations and data that are used in our analysis. In Section~\ref{SectionThree}, we present the nebular emission line properties of our intermediate-$z$ sample of galaxies. In Section~\ref{SectionFour}, we discuss the similarities and differences observed in the nebular spectra of galaxies at different redshifts spanning $z \simeq 0-2$. In Section~\ref{SectionFive}, we summarize our findings and their implications for studies of galaxy evolution during this period of cosmic time. Throughout this work, we report wavelengths in vacuum and adopt a standard flat $\Lambda$ cosmology with $\Omega_{\mathrm{m}} = 0.3$, $\Omega_{\Lambda} = 0.7$, and $H_{0} = 70\,\mathrm{km\,s^{-1}\,Mpc^{-1}}$. \par

%%%%%
\section{Observations and Data}
\label{SectionTwo}

\subsection{LEGA-C}
\label{LEGA-C}

LEGA-C is a European Southern Observatory (ESO) Public Spectroscopic Survey (PI: A. van der Wel, PID: 194-A.2005) of $N \approx 3500$ $K$-band selected galaxies at $0.6 < z < 1.0$ in the Cosmological Evolution Survey (COSMOS) field. The survey used the Visible Multi-Object Spectrograph \citep[VIMOS;][]{VIMOS} at the ESO VLT to obtain deep continuum spectra of a large sample of galaxies at intermediate redshift. These data have already been used to measure stellar population ages and star formation histories \citep{Chauke:2018, Wu:2018}, kinematics from stars and ionized gas \citep{Bezanson:2018a, Bezanson:2018b}, and molecular gas contents and scaling relations \citep{Spilker:2018}. This paper complements these efforts by investigating the emission line properties of the LEGA-C sample using measurements from Data Release 3 \citep[DR3;][]{LEGAC_DR3}. \par

\subsubsection{Sample Selection and Optical Spectroscopy}
\label{Sample}

The galaxies targeted by LEGA-C were selected from the UltraVISTA catalog on the basis of their photometric redshift and $K$-band magnitude. The $K$-band selection effectively produces a mass-representative sample at log($\mathrm{M_{\ast}/M_{\odot}}$) $\gtrsim 10.3$. Each galaxy was observed for $\sim 20$ hours using VIMOS, which produced spectra with typical signal-to-noise ratio $\mathrm{S/N} \sim 20$ \AA$^{-1}$ and resolution $\mathrm{R} \sim 3000$ across a wavelength range of $\lambda \sim 6300 - 8800\,$\AA. \par

We make use of the following quantities measured by the LEGA-C team for our analysis: spectroscopic redshifts ($z_{\mathrm{spec}}$), stellar masses (hereafter $\mathrm{M_{\ast}}$), star formation rates (SFRs), best-fit spectral energy distributions (SEDs), and emission line flux measurements (i.e., [O\,II]$\mathrm{\lambda\lambda 3727,3729}$; H$\gamma$; H$\beta$; and [O\,III]$\mathrm{\lambda\lambda 4960,5008}$). Spectroscopic redshifts were determined from stellar absorption features and measured by cross-correlating the observed spectra with a set of general templates. Stellar masses, SFRs, and best-fit SEDs were determined using photometric data from \citet{Muzzin:2013a, Muzzin:2013b} and Fitting and Assessment of Synthetic Templates \citep[FAST;][]{Kriek:2008}, adopting \citet{Bruzual:2003} stellar population libraries, a \citet{Chabrier:2003} initial mass function (IMF), the extinction curve from \citet{Calzetti:2000}, and exponentially declining star-formation histories. Emission line flux measurements were determined from best-fit model Gaussians after stellar continuum subtraction.

As we are concerned with the nebular properties of star-forming galaxies at $0.6 < z < 1.0$ from LEGA-C, we select all sources for which H$\beta$ and [O\,III]$\mathrm{\lambda 5008}$ are well-detected, with $\mathrm{S/N} > 3$ after continuum subtraction. This results in a sample of $N = 767$ star-forming galaxies, which is the primary sample that we consider throughout Section~\ref{SectionThreeOne}. The analysis presented in Sections~\ref{SectionThreeTwo} and \ref{SectionThreeThree} additionally requires $\mathrm{S/N} > 3$ measurements of [O\,II]$\mathrm{\lambda\lambda 3727,3729}$ and [O\,III]$\mathrm{\lambda 4960}$ after continuum subtraction, resulting in a secondary sample of $N = 192$ star-forming galaxies that have significant detections of [O\,II]$\mathrm{\lambda\lambda 3727,3729}$, H$\beta$, and [O\,III]$\mathrm{\lambda\lambda 4960,5008}$ (for this subsample, the signal-to-noise ratio requirement is for the sum of the lines that make up the doublet). All of the LEGA-C galaxies considered in this paper have [O\,II]$\mathrm{\lambda\lambda 3727,3729}$ line measurements from the LEGA-C DR3 catalogs. For galaxies at $z \gtrsim 0.75$, H$\beta$ and [O\,III]$\mathrm{\lambda\lambda 4960,5008}$ are shifted out of the optical spectra from VIMOS and instead must be recovered using NIR observations, as described below. \par 

\begin{figure}
    \centering
	\includegraphics[width=0.95\linewidth]{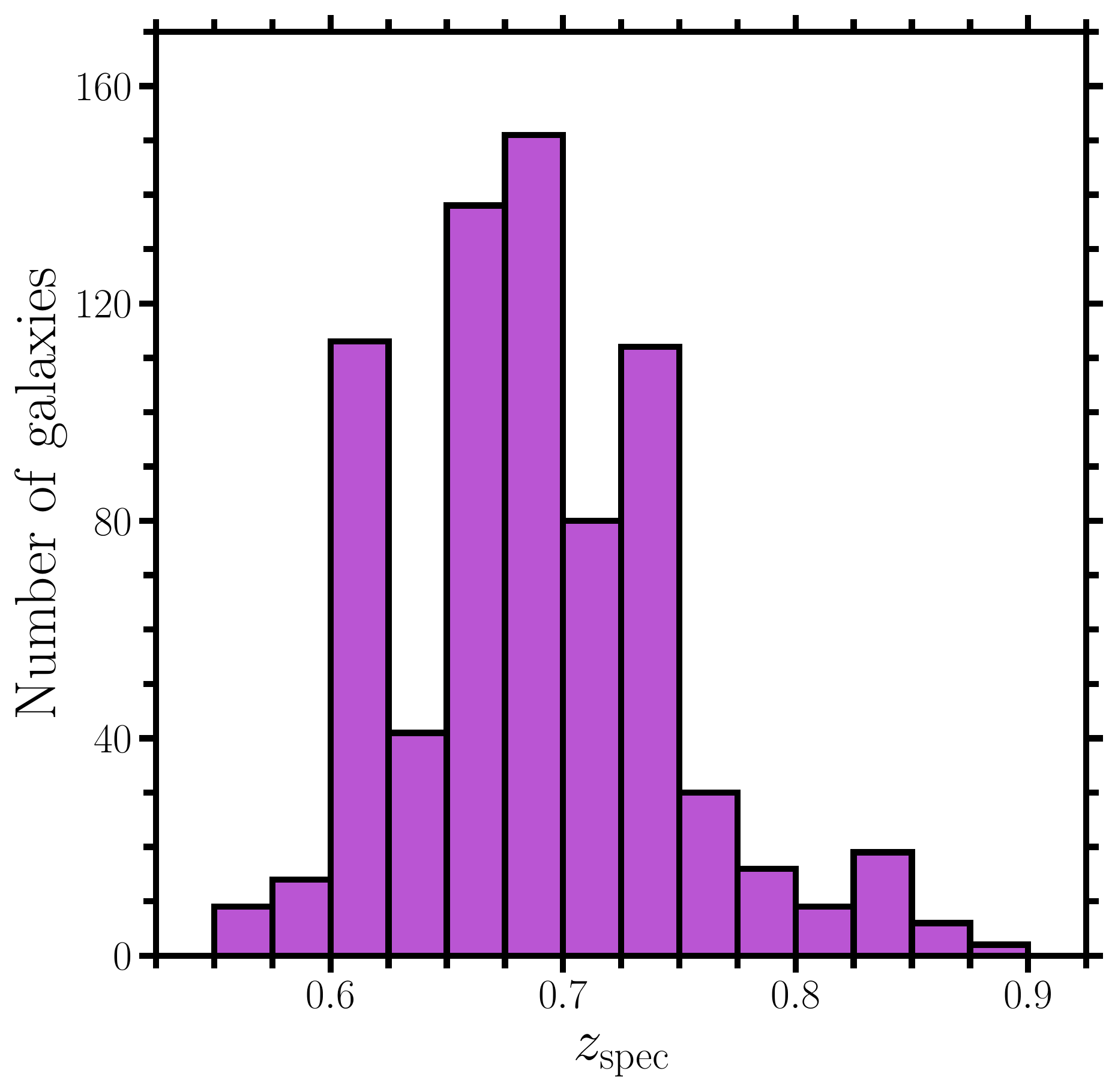}
    \caption{Histogram showing the distribution of spectroscopic galaxy redshifts for the primary sample of LEGA-C galaxies described in Section~\ref{Sample}. The median redshift value for the primary sample of LEGA-C galaxies is $z = 0.7$.}
    \label{fig:histogram_redshift}
\end{figure}

\begin{figure*}
    \centering
	\includegraphics[width=0.9\linewidth]{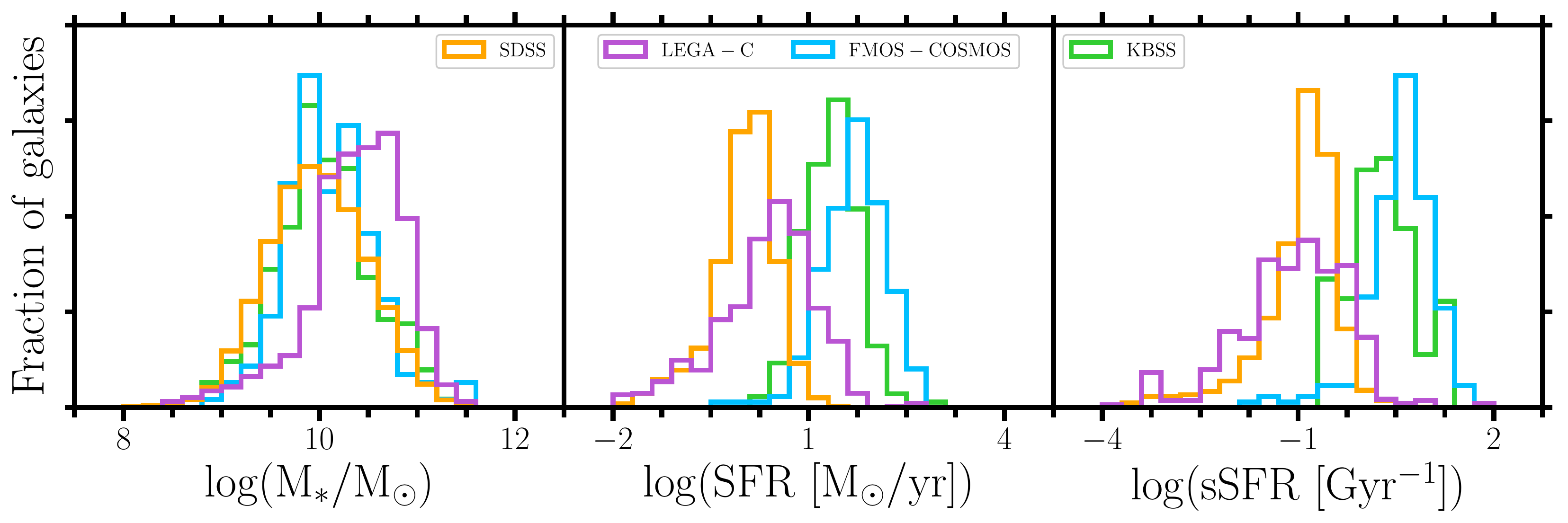}
    \caption{Histograms showing the distribution of galaxy properties ($\mathrm{M_{\ast}}$, SFR, and sSFR) for the primary sample of LEGA-C galaxies from Section~\ref{Sample} (shown in purple), the low-redshift comparison sample of SDSS galaxies from Section~\ref{SDSS} (shown in orange), the intermediate-redshift comparison sample of FMOS-COSMOS galaxies from Section~\ref{FMOS} (shown in blue), and the high-redshift comparison sample of KBSS galaxies from Section~\ref{KBSS} (shown in green).}
    \label{fig:histogram_properties}
\end{figure*}

Short descriptions of these samples are given in Table~\ref{tab:samples}, with the median redshift, $\mathrm{M_{\ast}}$, SFR, and specific star formation rate (sSFR = SFR/$\mathrm{M_{\ast}}$) provided in Table~\ref{tab:properties}. The median redshift of both  intermediate-redshift LEGA-C samples is $z = 0.7$. The median $\mathrm{M_{\ast}}$, SFR, and sSFR of the intermediate-redshift primary (secondary) sample of LEGA-C galaxies are $\mathrm{log}(\mathrm{M_{\ast}/M_{\odot}}) \approx 10.4$ ($10.3$), $\mathrm{log}(\mathrm{SFR}/[\mathrm{M_{\odot}/yr}]) \approx 0.4$ ($0.5$), and $\mathrm{log}(\mathrm{sSFR}/[\mathrm{Gyr}^{-1}]) \approx -1.0$ ($-0.8$). The redshift distribution of the intermediate-redshift primary sample of LEGA-C galaxies that we consider throughout Section~\ref{SectionThreeOne} is shown in purple in Figure~\ref{fig:histogram_redshift}. The $\mathrm{M_{\ast}}$, SFR, and sSFR distributions for the intermediate-redshift primary sample of LEGA-C galaxies that we consider throughout Section~\ref{SectionThreeOne} are shown in purple in Figure~\ref{fig:histogram_properties}. \par

% Histogram of Redshifts

% Table summarizing samples
\begin{center}
\begin{table*}
	\caption{Description of samples used throughout.} \label{tab:samples}
	\begin{threeparttable}
	\begin{tabular}{lll}
		\hline
		\hline
		Sample & Parent Survey & Selection Criteria \\
		\hline
		Low-redshift comparison & SDSS & $\mathrm{S/N} > 5$ for [O\,II]$\mathrm{\lambda\lambda 3727,3729}$; H$\beta$;  [O\,III]$\mathrm{\lambda\lambda 4960,5008}$; \\
		& & [N\,II]$\mathrm{\lambda 6585}$; H$\alpha$ \\
		Low-redshift, mass-matched & SDSS & $\mathrm{S/N} > 5$ for [O\,II]$\mathrm{\lambda\lambda 3727,3729}$; H$\beta$;  [O\,III]$\mathrm{\lambda\lambda 4960,5008}$; \\
		& & [N\,II]$\mathrm{\lambda 6585}$; H$\alpha$ \\
		Intermediate-redshift primary & LEGA-C & $\mathrm{S/N} > 3$ for H$\beta$; [O\,III]$\mathrm{\lambda 5008}$ \\
		Intermediate-redshift secondary & LEGA-C & $\mathrm{S/N} > 3$ for [O\,II]$\mathrm{\lambda\lambda 3727,3729}$; H$\beta$; [O\,III]$\mathrm{\lambda\lambda 4960,5008}$ \\
		Intermediate-redshift comparison & FMOS-COSMOS & $\mathrm{S/N} > 3$ for H$\beta$; [O\,III]$\mathrm{\lambda 5008}$ \\
		High-redshift comparison & KBSS & $\mathrm{S/N} > 3$ for [O\,II]$\mathrm{\lambda\lambda 3727,3729}$; H$\beta$; [O\,III]$\mathrm{\lambda\lambda 4960,5008}$; \\
		& & [N\,II]$\mathrm{\lambda 6585}$; H$\alpha$ \\
		\hline \\
	\end{tabular}
	\end{threeparttable}
\end{table*}
\end{center}

% Table summarizing galaxy properties percentiles
\begin{center}
\begin{table*}
	\caption{Description of properties of the samples used throughout. Median values are reported for physical quantities.} \label{tab:properties}
	\begin{threeparttable}
	\begin{tabular}{lccccc}
		\hline
		\hline
		Sample & $N_{\mathrm{gal}}$ & $z$ & $\mathrm{log}(\mathrm{M_{\ast}/M_{\odot}})$ & $\mathrm{log}(\mathrm{SFR}/[\mathrm{M_{\odot}/yr}])$ & $\mathrm{log}(\mathrm{sSFR}/[\mathrm{Gyr}^{-1}])$ \\
		\hline
        Low-redshift comparison & 116013 & $0.1$ & $10.01$ & $+0.069$ & $-0.878$ \\
		Low-redshift, mass-matched & 7670 & $0.1$ & $10.42$ & $+0.142$ & $-1.164$ \\
		Intermediate-redshift primary & 767 & $0.7$ & $10.43$ & $+0.430$ & $-1.000$ \\
		Intermediate-redshift secondary & 192 & $0.7$ & $10.31$ & $+0.500$ & $-0.810$ \\
		Intermediate-redshift comparison & 192 & $1.6$ & $10.07$ & $+1.701$ & $+0.624$ \\
		High-redshift comparison & 380 & $2.3$ & $10.05$ & $+1.332$ & $+0.271$ \\
		\hline
	\end{tabular}
	\end{threeparttable}
\end{table*}
\end{center}

% Histogram of Galaxy Properties

\vspace{-16mm}
\subsubsection{Near-infrared Spectroscopy}
\label{Subsample}

NIR spectroscopic observations were conducted for a subsample of LEGA-C galaxies to enable measurements of emission lines at red rest-optical wavelengths that were missing from the optical VIMOS spectra. Galaxies from LEGA-C were prioritized for follow-up if they had $\mathrm{S/N} > 3$ measurements of [O\,II]$\mathrm{\lambda\lambda 3727,3729}$ from VIMOS, in order to maximize the likelihood of recovering other optical emission lines. \par

\begin{figure*}
    \centering
	\includegraphics[width=0.9\textwidth]{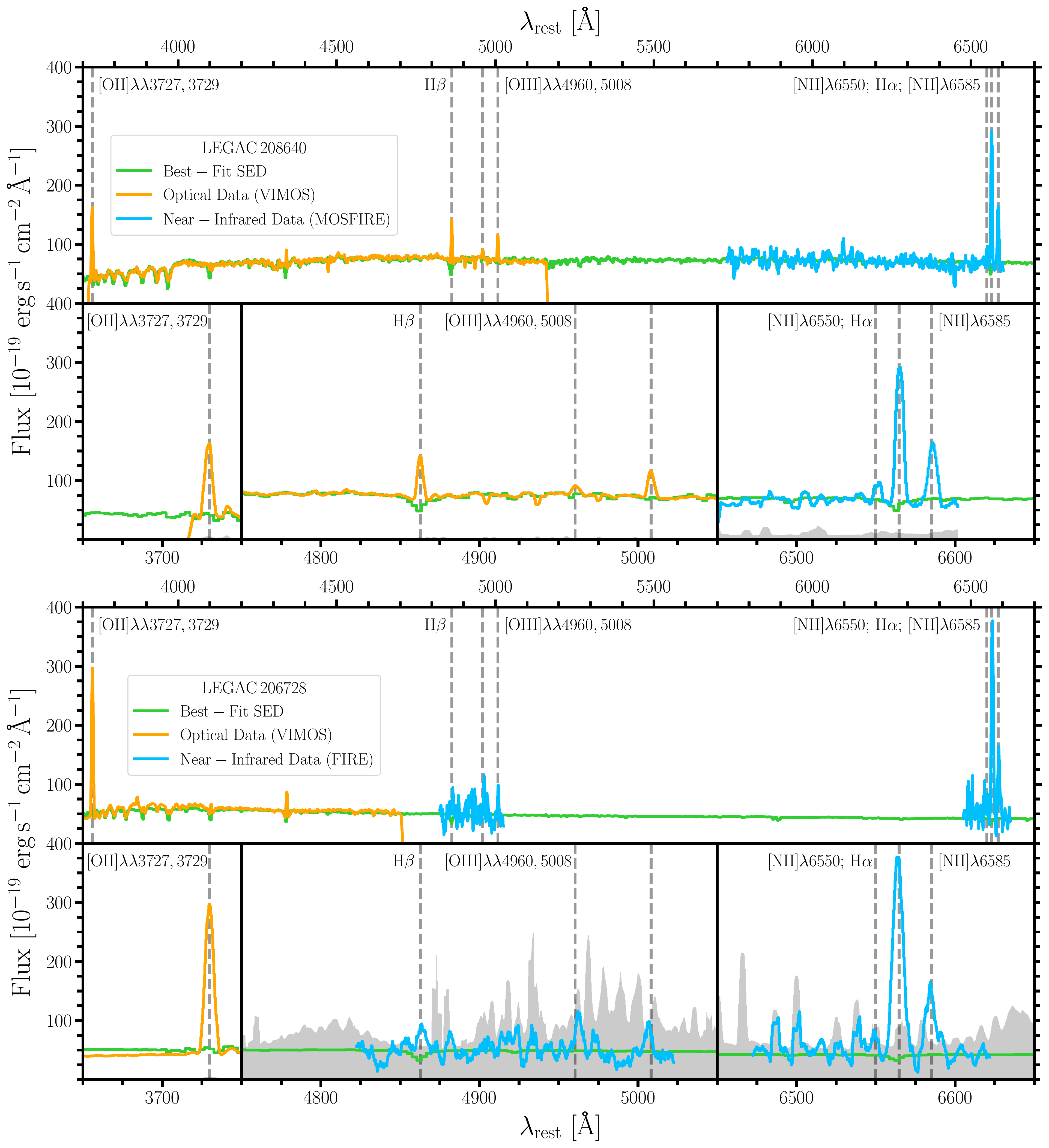}
    \caption{Shown in the upper panels, VIMOS+MOSFIRE spectra of one of the LEGA-C galaxies within our subsample described in Section~\ref{Subsample} (ID: 208640, $z \sim 0.7$). Shown in the lower panels, VIMOS+FIRE spectra from one of the LEGA-C galaxies within our subsample described in Section~\ref{Subsample} (ID: 206728, $z \sim 0.9$). The orange line represents the optical data from VIMOS, the blue line represents the near-infrared data from MOSFIRE or FIRE, and the green line represents the best-fit SED model from the LEGA-C team (i.e., the best-fit combination of stellar population templates). The regions surrounding prominent emission features ([O\,II]$\mathrm{\lambda\lambda 3727,3729}$, H$\beta$, [O\,III]$\mathrm{\lambda\lambda 4960,5008}$, [N\,II]$\mathrm{\lambda\lambda 6550,6585}$, and H$\alpha$) are shown with zoomed-in panels, with errors represented by the shaded grey regions. In the upper panels, these emission lines are all detected at $\mathrm{S/N} > 3$ from the VIMOS+MOSFIRE spectra. In the upper panels, these emission lines are all detected at $\mathrm{S/N} > 3$ from the VIMOS+FIRE spectra except for [O\,III]$\mathrm{\lambda\lambda 4960,5008}$. This figure illustrates how galaxies at intermediate redshifts currently require both observed-optical and observed-NIR spectra to offer complete coverage of the rest-optical spectrum.}
    \label{fig:spectra}
\end{figure*}

For $N = 43$ galaxies at $0.5 \lesssim z \lesssim 0.8$, these NIR data were obtained with the Multi-Object Spectrometer For InfraRed Exploration \citep[MOSFIRE;][]{McLean:2010, McLean:2012} on the Keck I Telescope. Observations were made using the $Y$-band or $J2$-band filters over two observing runs in March 2017 and November 2020. The $J2$-band filter can access a specific wavelength region that the $J$-band filter does not cover. These spectra have $\mathrm{R} \sim 3400$ and were acquired using 0.7 arcsec slits with a two-point AB dither ``mask nod'' pattern and a nod amplitude of 3.0 arcsec. Individual integration times are $120\,\mathrm{s}$ in both $Y$ and $J2$, read out with 16 read pairs. The total integration time for an individual galaxy is $\approx 3600\,\mathrm{s}$. The data were reduced using the publicly available data reduction pipeline\footnote{\url{https://www2.keck.hawaii.edu/inst/mosfire/drp.html}}. The resulting two-dimensional (2D) spectra are then flux-calibrated, corrected for telluric absorption, and shifted to account for heliocentric velocity at the time of observation. One-dimensional (1D) spectra are extracted using boxcar apertures. At $0.5 < z < 0.8$, the LEGA-C galaxies are more extended than a point source. To account for slit losses and place these 1D spectra on the same flux scale as the optical spectra from VIMOS, the continuum near bright emission lines is scaled to match the best-fit SED model (see Figure~\ref{fig:spectra}). \par

An additional $N = 10$ LEGA-C galaxies at $0.8 \lesssim z \lesssim 1.0$ were observed using the Folded-Port Infrared Echellette \citep[FIRE;][]{Simcoe:2008, Simcoe:2010} on the Magellan Baade Telescope. Because FIRE is a cross-dispersed echellette, it can observe portions of the NIR that are inaccessible to MOSFIRE, which uses order-sorting filters. In addition to the prioritization based on $\mathrm{S/N} > 3$ measurements of [O\,II]$\mathrm{\lambda\lambda 3727,3729}$ from VIMOS, galaxies were also prioritized by their $J$ magnitude, which ranges from $J \approx 20.2 - 21.2 $ for the observed subsample. At these magnitudes, blind offsets can be avoided and instead targets can be directly acquired using the FIRE slit-viewing camera. Observations were made over two observing runs in February 2019 and February 2020. These spectra have $\mathrm{R} \sim 6000 - 8000$ and were acquired using 1.0 arcsec slits with a two-point AB dither ``mask nod'' pattern and a nod amplitude of 2.5 arcsec. The integration times for individual exposures are $\approx 1200 -1800\,\mathrm{s}$. The initial integration time for an individual galaxy was $\approx 3600\,\mathrm{s}$, although some galaxies were targeted for additional follow-up observations in order to recover [N\,II], which is typically the faintest emission line of interest in the observed-NIR. The data were reduced using the publicly-available data reduction pipeline, \texttt{FIREHOSE}\footnote{\url{http://web.mit.edu/rsimcoe/www/FIRE/ob_data.htm}}, with boxcar extraction apertures defined interactively based on the extent of the H$\alpha$ line in individual frames. \texttt{FIREHOSE} performs a first-order flux calibration and telluric corrections using observations of a bright standard star taken immediately before or after the science target. As with the MOSFIRE observations, the continuum in the resulting 1D spectra is scaled to match the SED model continuum (see Figure~\ref{fig:spectra}). \par

For each of the $N = 53$ galaxies with observations from MOSFIRE or FIRE, we measured line fluxes, redshifts, and velocity dispersions using a spectral modeling technique where the best-fit SED  is adopted as the continuum, and emission lines are fit by a set of Gaussian line profiles with a single redshift and velocity dispersion. This method implicitly accounts for Balmer absorption features. We performed non-linear least-squares minimization using \texttt{LMFIT} \citep{LMFIT} optimization techniques and estimated standard errors on the inferred model parameters from the covariance matrix. Examples of the observed VIMOS+MOSFIRE spectra and the VIMOS+FIRE spectra and the corresponding best-fit spectral models are shown in Figure~\ref{fig:spectra}. There are $N = 16$ galaxies with observations from MOSFIRE or FIRE that have at least one emission line with $\mathrm{S/N} > 3$. These galaxies are included in the analysis presented in Section~\ref{SectionThree} and the discussion presented in Section~\ref{SectionFour}. \par

% Spectra

\subsection{Comparison Samples}
\label{Comparison}

\subsubsection{Low-redshift Comparison Sample}
\label{SDSS}

The low-redshift comparison sample of galaxies is built from Sloan Digital Sky Survey (SDSS) DR8 \citep{Gunn:2006, Eisenstein:2011, SDSS:DR8}. We select galaxies that are primary targets (SCIENCEPRIMARY == 1), safe to use scientifically (USE == 1), and fall within a specific redshift range ($0.04 < z < 0.1$). The lower redshift cut is made to avoid strong aperture effects. We also select galaxies for which [O\,II]$\mathrm{\lambda\lambda 3727,3729}$; H$\beta$; [O\,III]$\mathrm{\lambda\lambda 4960,5008}$; [N\,II]$\mathrm{\lambda 6585}$; and H$\alpha$ have been well-detected, with signal-to-noise ratios $\mathrm{S/N} > 5$ for each of the individual emission lines (for emission lines that are part of a doublet, the signal-to-noise ratio requirement is for the sum of the lines that make up the doublet). \par

We ultimately obtain a sample of $N = 116013$ galaxies from SDSS DR8. Emission line fluxes and $\mathrm{M_{\ast}}$ were obtained from the MPA-JHU Value Added Catalogs provided by the Max-Planck Institute for Astronomy and John Hopkins University, following the methodology described by \citet{Tremonti:2004}. Stellar mass is reported as the median of the probability density functions for the ``total'' values, using the Bayesian methodology and model grids described in \citet{Kauffmann:2003} which use a \citet{Kroupa:2001} IMF. To convert $\mathrm{M_{\ast}}$ from a \citet{Kroupa:2001} IMF to a \citet{Chabrier:2003} IMF, we multiply by a constant factor 0.92 which corresponds to a change in $\mathrm{log}(\mathrm{M_{\ast}/M_{\odot}})$ of $-$0.034 dex \citep[][]{Madau:2014}. For a more detailed description of these catalogs, we refer the reader to Section~4.3.2 of \citet{SDSS:DR8}.\par

A short description of this sample is given in Table~\ref{tab:samples} while redshift, $\mathrm{M_{\ast}}$, SFR, and sSFR median values are given in Table~\ref{tab:properties}. The median $\mathrm{M_{\ast}}$, SFR, and sSFR of the low-redshift comparison sample are $\mathrm{log}(\mathrm{M_{\ast}/M_{\odot}}) \approx 10.0$, $\mathrm{log}(\mathrm{SFR}/[\mathrm{M_{\odot}/yr}]) \approx 0.1$, and $\mathrm{log}(\mathrm{sSFR}/[\mathrm{Gyr}^{-1}]) \approx -0.9$. The $\mathrm{M_{\ast}}$, SFR, and sSFR distributions for the low-redshift comparison sample that we consider throughout Sections~\ref{SectionThree} and \ref{SectionFour} are shown in orange in Figure~\ref{fig:histogram_properties}. \par

\subsubsection{Intermediate-redshift Comparison Sample}
\label{FMOS}

The intermediate-redshift comparison sample of galaxies is built from the Fiber Multi-Object Spectrograph (FMOS)-COSMOS Survey \citep[][]{Silverman:2015, Kashino:2019}. FMOS-COSMOS is a large rest-optical spectroscopic survey of star-forming galaxies at $z \sim 1.6$. This survey was conducted over the $1.7\,\mathrm{deg}^{2}$ COSMOS field using the FMOS instrument on the Subaru Telescope. Similar to LEGA-C, the parent sample of FMOS-COSMOS galaxies was selected from the UltraVISTA catalog on the basis of their photometric redshift and $K$-band magnitude. We select all sources for which H$\beta$ and [O\,III]$\lambda5008$ have been well-detected, with signal-to-noise ratios $\mathrm{S/N} > 3$ for each emission line. \par

We ultimately obtain a sample of $N = 192$ galaxies from FMOS-COSMOS. Emission line fluxes were measured using the Interactive Data Language (IDL) routine \texttt{MPFIT} \citep[][]{MPFIT} as described in \citet{Kashino:2019}. Stellar mass was measured from reddened stellar population synthesis models that were fit to broadband photometry, following the methodology and model grids described in \citet[][]{Ilbert:2015} which use a \citet{Chabrier:2003} IMF. For a more detailed description of these measurements of $\mathrm{M_{\ast}}$, we refer the reader to Section~4 of \citet{Laigle:2016}. \par

A short description of this sample is given in Table~\ref{tab:samples} while redshift, $\mathrm{M_{\ast}}$, SFR, and sSFR median values are given in Table~\ref{tab:properties}. The median $\mathrm{M_{\ast}}$, SFR, and sSFR of the high-redshift comparison sample are $\mathrm{log}(\mathrm{M_{\ast}/M_{\odot}}) \approx 10.1$, $\mathrm{log}(\mathrm{SFR}/[\mathrm{M_{\odot}/yr}]) \approx 1.7$, and $\mathrm{log}(\mathrm{sSFR}/[\mathrm{Gyr}^{-1}]) \approx 0.6$. The $\mathrm{M_{\ast}}$, SFR, and sSFR distributions for the intermediate-redshift comparison sample that we consider throughout Sections~\ref{SectionThree} and \ref{SectionFour} are shown in blue in Figure~\ref{fig:histogram_properties}. \par

\subsubsection{High-redshift Comparison Sample}
\label{KBSS}

The high-redshift comparison sample of galaxies is built from the MOSFIRE component of the Keck Baryonic Structure Survey \citep[KBSS;][]{Rudie:2012, Steidel:2014}. KBSS is a large spectroscopic survey of star-forming galaxies at $2.0 \lesssim z \lesssim 2.6$. This survey was conducted in 15 separate survey fields with a total survey area of $0.24\,\mathrm{deg}^{2}$. In contrast with LEGA-C and FMOS-COSMOS, which employ a rest-optical magnitude limit, the parent sample of KBSS galaxies was selected on the basis of their rest-ultraviolet (UV) colors \citep{Adelberger:2004, Steidel:2004}. However, this does not significantly impact the comparison between KBSS and LEGA-C galaxies at $\mathrm{log}(\mathrm{M_{\ast}/M_{\odot}}) \gtrsim 10$, where KBSS includes galaxies with SFRs down to $\approx 5\ \mathrm{M_{\odot}}\,\mathrm{yr}^{-1}$, well below the star-forming main sequence even at these masses \citep[e.g.,][]{Whitaker:2017}. Despite differences in galaxy selection technique, the demographics and nebular properties of $z \sim 2$ KBSS galaxies are consistent with $z \sim 2$ galaxy samples selected using a combination of photometric redshift and rest-optical magnitude, more similar to LEGA-C and FMOS-COSMOS \citep[][]{Runco:2022}. \par

Substantial multiwavelength imaging, rest-UV spectroscopy, and rest-optical spectroscopy have been conducted in all KBSS fields. The photometry, spectra, and derived quantities are described in detail elsewhere \citep{Steidel:2003, Reddy:2012, Steidel:2014, Strom:2017}. Here, we summarize the most relevant details of the rest-optical spectroscopic observations. \par

At $2.0 \lesssim z \lesssim 2.6$, the emission features of interest ([O\,II]$\mathrm{\lambda\lambda 3727,3729}$; H$\beta$; [O\,III]$\lambda\lambda 4960,5008$; [N\,II]$\mathrm{\lambda 6585}$; and H$\alpha$) are accessible in the $J$-, $H$-, and $K$-bands. Observations were made using MOSFIRE with typical total exposure times of approximately $7200 - 14400\,\mathrm{s}$ per band, or until the strong lines are detected at $\mathrm{S/N} > 3$. The data were reduced and analyzed in the same way as the LEGA-C galaxies observed with MOSFIRE from Section~\ref{Subsample}. Slit losses were determined as described in \citet{Strom:2017}. We select all sources for which [O\,II]$\mathrm{\lambda\lambda 3727,3729}$; H$\beta$; [O\,III]$\mathrm{\lambda\lambda 4960,5008}$; [N\,II]$\mathrm{\lambda 6585}$; and H$\alpha$ have been well-detected, with signal-to-noise ratios $\mathrm{S/N} > 3$ for each of the individual emission lines (for emission lines that are part of a doublet, the signal-to-noise ratio requirement is for each of the sum of the lines that make up the doublet). \par

We ultimately obtain a sample of $N = 380$ galaxies from KBSS. Emission line fluxes were measured using the custom IDL\footnote{Interactive Data Language (IDL).} routine \texttt{mospec}\footnote{\url{https://github.com/allisonstrom/mospec}}, as described in \citet{Strom:2017}. Stellar mass was measured from reddened stellar population synthesis models that were fit to broadband photometry, following the methodology and model grids described in \citet[][]{Reddy:2012} and \citet[][]{Steidel:2014} which use a \citet{Chabrier:2003} IMF. For a more detailed description of these measurements, we refer the reader to Section~\ref{SectionTwo} of \citet{Strom:2017}. \par

A short description of this sample is given in Table~\ref{tab:samples} while redshift, $\mathrm{M_{\ast}}$, SFR, and sSFR median values are given in Table~\ref{tab:properties}. The median $\mathrm{M_{\ast}}$, SFR, and sSFR of the high-redshift comparison sample are $\mathrm{log}(\mathrm{M_{\ast}/M_{\odot}}) \approx 10.1$, $\mathrm{log}(\mathrm{SFR}/[\mathrm{M_{\odot}/yr}]) \approx 1.3$, and $\mathrm{log}(\mathrm{sSFR}/[\mathrm{Gyr}^{-1}]) \approx 0.3$. The $\mathrm{M_{\ast}}$, SFR, and sSFR distributions for the high-redshift comparison sample that we consider throughout Sections~\ref{SectionThree} and \ref{SectionFour} are shown in green in Figure~\ref{fig:histogram_properties}. \par

%%%%%
\section{Analysis}
\label{SectionThree}

\subsection{The Mass-Excitation Diagram}
\label{SectionThreeOne}

The nebular spectra of galaxies provide useful diagnostics for a number of physical conditions in galaxies, including enrichment and ionization. Perhaps the most famous example is the diagram comparing [O\,III]$\lambda5008$/H$\beta$ (hereafter O3; see Table~\ref{tab:definitions}) and [N\,II]$\lambda6585$/H$\alpha$ (hereafter N2; see Table~\ref{tab:definitions}), which was one of three line ratio diagrams popularized by \citet{Veilleux:1987}. This ``BPT diagram'' \citep[after Baldwin$-$Phillips$-$Terlevich;][]{Baldwin:1981} is frequently used to determine the dominant ionizing source in $z \sim 0$ galaxies and to separate star-forming galaxies from AGN. Other parameters also affect where objects fall on this diagram: galaxies with lower mass and higher ionization are typically found towards the upper left portion of the BPT diagram (with higher values of O3 and lower values of N2), while galaxies with higher mass and lower ionization are typically found towards the lower right portion of the BPT diagram (with lower values of O3 and higher values of N2). However, the BPT diagram requires observations spanning a relatively wide range in wavelength and cannot be used for $z \gtrsim 0.5$ galaxies with only observed-optical spectra, as [N\,II]$\lambda6585$ and H$\alpha$ will be redshifted to NIR wavelengths.\par

\citet{Juneau:2011} introduced the MEx (comparing O3 and $\mathrm{M_{\ast}}$) as an alternative to the BPT diagram for galaxies lacking measurements of N2 (such as galaxies at intermediate redshifts, $0.6 < z < 1.0$). For $z \sim 0$ galaxies, there is a strong anti-correlation observed between O3 and $\mathrm{M}_{\ast}$, reflecting an overall increase in oxygen abundance and decrease in ionization and excitation with increasing $\mathrm{M}_{\ast}$. At high $\mathrm{M}_{\ast}$, AGN separate from star-forming galaxies in the MEx because they have harder ionizing spectra than massive stars \citep{Veilleux:1987}: galaxies at high $\mathrm{M}_{\ast}$ would generally have low values of O3 if not for the presence of an AGN. For $z \sim 1.6$ and $z \sim 2$ galaxies, a strong anti-correlation is also observed between O3 and $\mathrm{M}_{\ast}$, but these galaxies are significantly offset towards higher values of O3 when compared to $z \sim 0$ galaxies, even in the absence of an AGN. This significant offset is instead the result of a combination of differences in the properties of stellar populations \citep[e.g., shape of the ionizing radiation field;][]{Steidel:2016, Strom:2017, Shapley:2019, Topping:2020, Runco:2021} and the properties of $\mathrm{H\,II}$ regions surrounding massive stars \citep[e.g., gas-phase metallicity;][]{Maiolino:2008}. \par

% Table summarizing strong-line ratios
\begin{center}
\begin{table}
	\centering
    \caption{Definitions of strong line indices.} \label{tab:definitions}
	\begin{tabular}{lccccc}
		\hline
		\hline
		Index & Definition \\
		\hline
		N2 & log([N\,II]$\mathrm{\lambda 6585}$/H$\alpha$) \\
		O3 & log([O\,III]$\lambda5008$/H$\beta$) \\
		O32 & log([O\,III]$\mathrm{\lambda\lambda 4960,5008}$/[O\,II]$\mathrm{\lambda\lambda 3727,3729}$) \\
		R23 & log(([O\,II]$\mathrm{\lambda\lambda 3727,3729}$ + [O\,III]$\mathrm{\lambda\lambda 4960,5008}$)/H$\beta$) \\
		N2O2 & log([N\,II]$\lambda$6585/[O\,II]$\lambda\lambda$3727,3729) \\
		\hline
	\end{tabular}
\end{table}
\end{center}

% Mass-Excitation Diagram
\begin{figure*}
    \centering
	\includegraphics[width=0.7\linewidth]{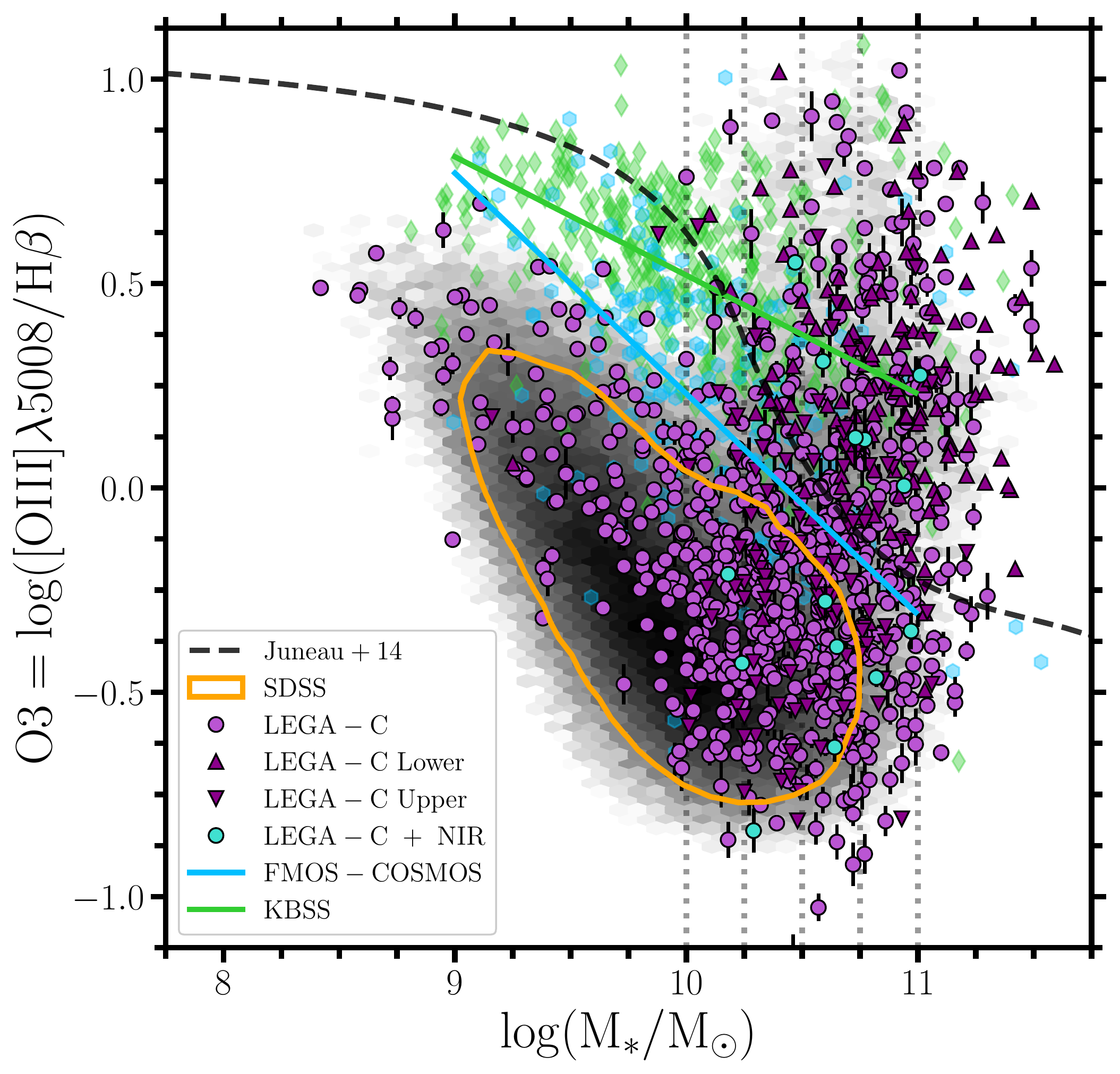}
    \caption{The stellar mass-excitation diagram (MEx). The primary sample of $0.6 \lesssim z \lesssim 1.0$ LEGA-C galaxies from Section~\ref{Sample} is shown with the purple circles. The subsample of LEGA-C galaxies from Section~\ref{Subsample} with observed-NIR spectra from MOSFIRE or FIRE are shown with the cyan circles. LEGA-C galaxies with $3\sigma$ lower (upper) limits on O3 are shown with the dark purple upward-facing (downward-facing) triangles. The low-redshift comparison sample of $z \sim 0$ SDSS galaxies from Section~\ref{SDSS} is shown in grayscale with an orange contour, where the contour encloses roughly 90\% of the low-redshift sample. The intermediate-redshift comparison sample of $z \sim 1.6$ FMOS-COSMOS galaxies from Section~\ref{FMOS} is shown with the blue hexagons and blue line, where the line represents the best-fit linear relation to the FMOS-COSMOS sample. The high-redshift comparison sample of $z \sim 2$ KBSS galaxies from Section~\ref{KBSS} is shown with the green diamonds and green line, where the line represents the best-fit linear relation to the KBSS sample. The division between $z = 0.7$ star-forming/composite galaxies and AGN from \citet{Juneau:2014} is given by the black dashed line. The black dotted vertical lines indicate the $\mathrm{M}_{\ast}$ bins that are used throughout. We see that the $0.6 \lesssim z \lesssim 1.0$ LEGA-C galaxies appear much more similar to the $z \sim 0$ SDSS galaxies than the $z \sim 1.6$ FMOS-COSMOS galaxies and the $z \sim 2$ KBSS galaxies.}
    \label{fig:MEX_O3}
\end{figure*}

Figure~\ref{fig:MEX_O3} shows the MEx for $0.6 \lesssim z \lesssim 1.0$ galaxies from LEGA-C. The primary sample of $0.6 \lesssim z \lesssim 1.0$ LEGA-C galaxies from Section~\ref{Sample} is shown with the purple circles. The subsample of LEGA-C galaxies from Section~\ref{Subsample} with observed-NIR spectra are shown with the cyan circles. LEGA-C galaxies with $3\sigma$ lower (upper) limits on O3 are shown with the dark purple upward-facing (downward-facing) triangles. The sample with lower limits on O3 consists of $N = 129$ galaxies that have [O\,III]$\lambda5008$ measurements with $\mathrm{S/N} > 3$ but H$\beta$ measurements with $\mathrm{S/N} < 3$. The sample with upper limits on O3 consists of $N = 112$ galaxies that have H$\beta$ measurements with $\mathrm{S/N} > 3$ but [O\,III]$\lambda5008$ measurements with $\mathrm{S/N} < 3$. Characteristic errors on the $\mathrm{M}_{\ast}$ from LEGA-C are $\simeq 0.1 - 0.2$~dex \citep{Muzzin:2013a, Muzzin:2013b}. \par

To place these results in context with galaxies that were forming at other times, we compare the LEGA-C sample with the three samples described in Section~\ref{Comparison}. The $z \sim 0$ SDSS galaxies are shown in grayscale with an orange contour, where the contour encloses roughly 90\% of the low-redshift sample. The individual $z \sim 1.6$ FMOS-COSMOS galaxies are shown with the blue hexagons and blue line, where the line represents the best-fit linear relation to the FMOS-COSMOS sample reported by \citet[][]{Kashino:2019}. The $z \sim 2$ KBSS galaxies are shown with the green diamonds and green line, where the line represents the best-fit linear relation to the KBSS sample from \citet[][]{Strom:2017}. \par

As shown in prior work \citep{Juneau:2011, Coil:2015, Strom:2017, Kashino:2019}, galaxies at all redshifts exhibit a sharp decline in O3 with increasing $\mathrm{M}_{\ast}$. This is particularly apparent at low $\mathrm{M_{\ast}}$ for the LEGA-C sample, which has a lower envelope that largely mirrors the behavior of SDSS. In contrast, the FMOS-COSMOS and KBSS galaxies are almost entirely disjoint with respect to the SDSS galaxies, with significantly elevated O3 at fixed $\mathrm{M}_{\ast}$. These offsets are likely the result of differences in gas-phase oxygen abundance and/or the typical shape of the ionizing radiation field. \par

The black dashed line in Figure~\ref{fig:MEX_O3} indicates the curve \citet{Juneau:2014} proposed to divide star-forming/composite galaxies from AGN at $z = 0.7$, which is offset from the analogous curve at $z \sim 0$ because galaxy samples at intermediate or high redshift are typically subject to emission line detection limits. This empirical division was calibrated using $z \sim 0$ galaxies from SDSS \citep[][]{Abazajian:2009}, as well as $z \sim 1$ galaxies from the Team Keck Treasury Redshift Survey \citep[TKRS][]{Wirth:2004} and the DEEP2 Galaxy Evolution Survey \citep[][]{Davis:2003, Newman:2013}. Accordingly, this empirical division should also apply to $0.6 \lesssim z \lesssim 1.0$ LEGA-C galaxies; we return to the issue of AGN contamination in Section~\ref{SectionFourTwo}. \par

% Histogram of Masses
\begin{figure}
    \centering
	\includegraphics[width=0.95\linewidth]{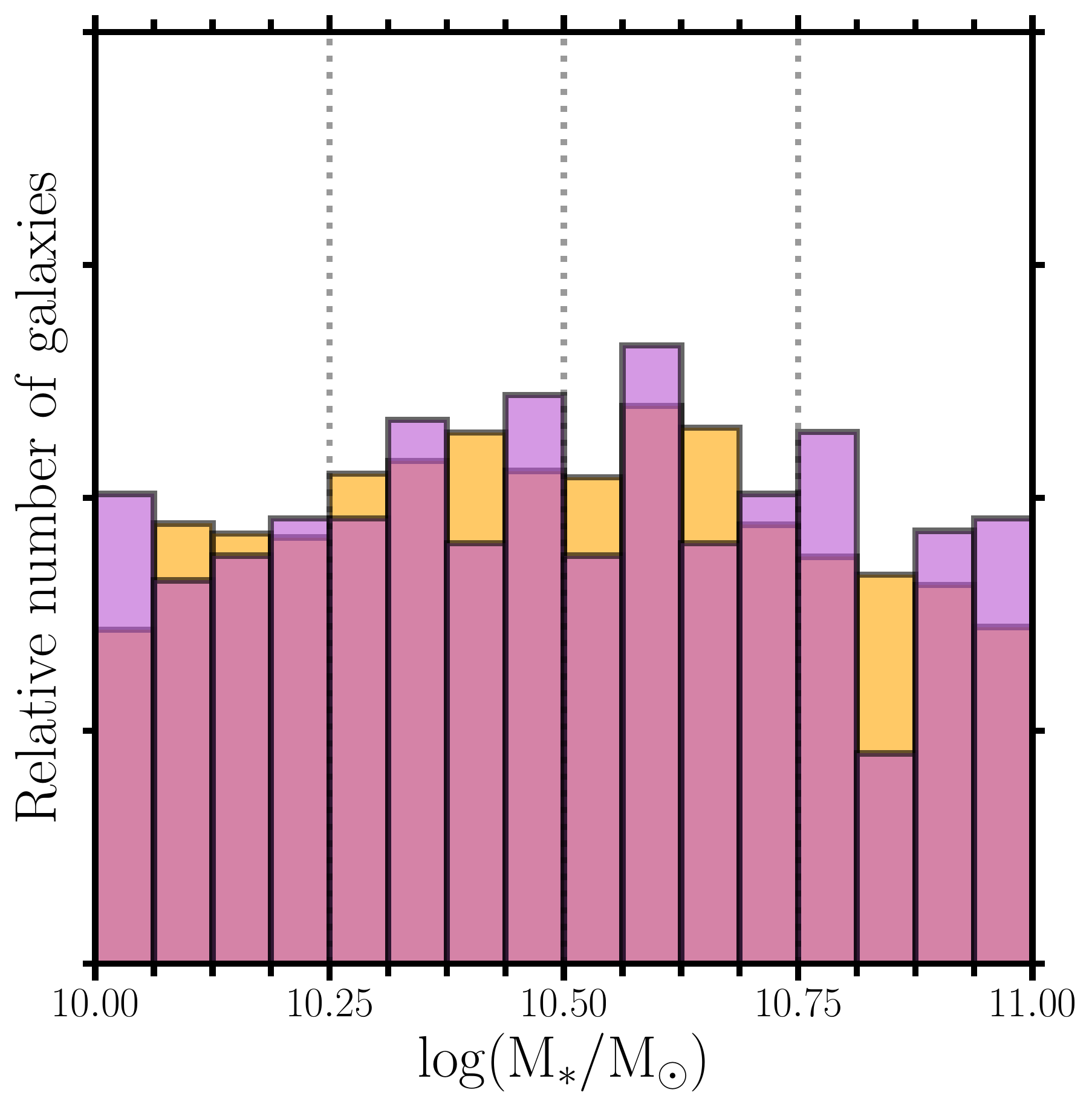}
    \caption{Histograms showing the distribution of $\mathrm{M}_{\ast}$ for galaxies within the primary sample of LEGA-C galaxies from Section~\ref{Sample} (shown in purple) and the low-redshift mass-matched comparison sample of $z \sim 0$ SDSS galaxies from Section~\ref{SectionThreeOne} (shown in orange). The black dotted lines indicate the $\mathrm{M}_{\ast}$ bins that are used throughout. The results of KS and AD tests indicate that these samples are consistent with being drawn from the same parent population in each of the $\mathrm{M}_{\ast}$ bins shown here, with $p_{\mathrm{KS}} \gtrsim 0.95$ for each of the individual bins.}
    \label{fig:histogram_mass}
\end{figure}

% [O\,III]/Hbeta CDF in bins of stellar mass
\begin{figure*}
    \centering
    \includegraphics[width=0.7\linewidth]{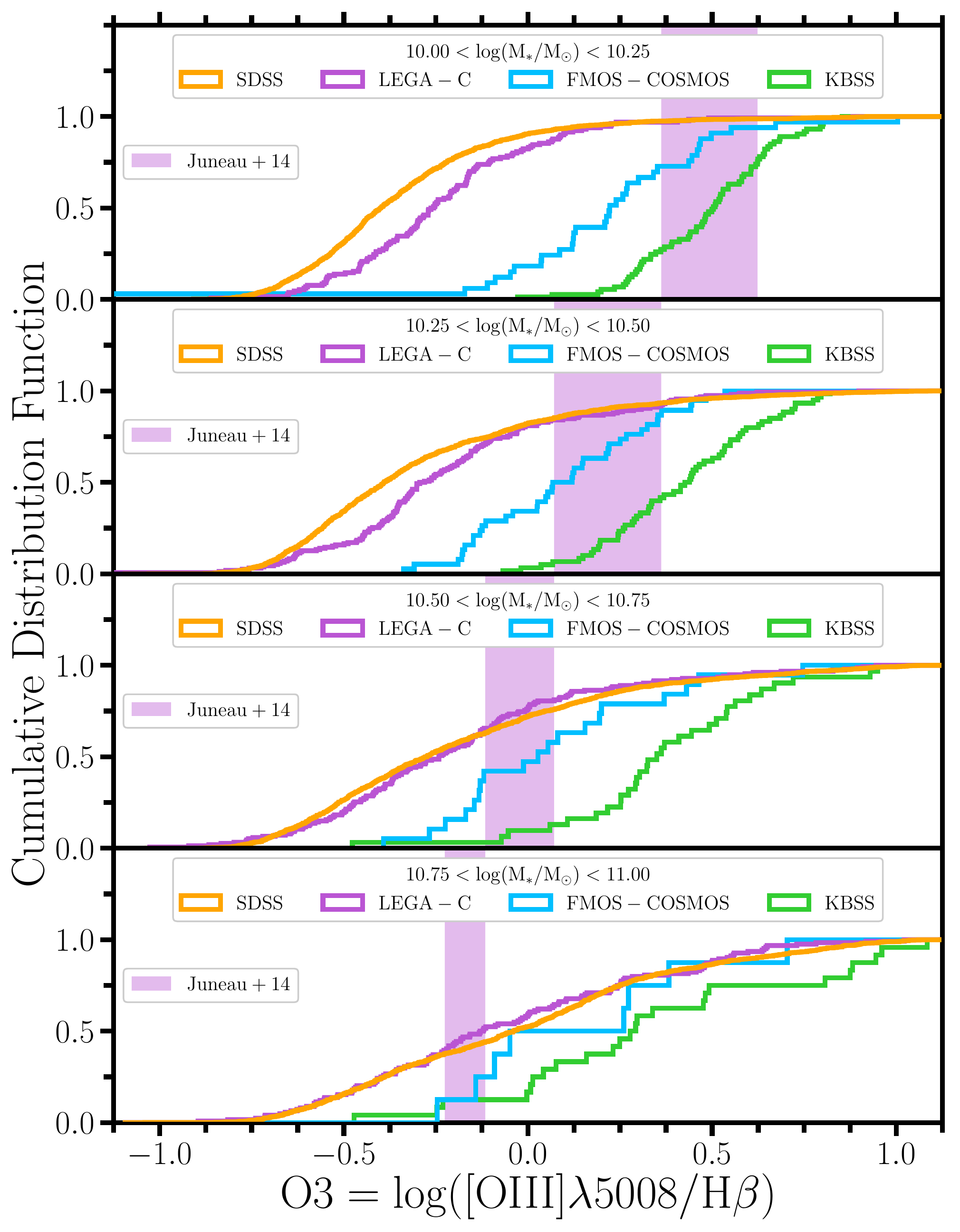}
    \caption{The cumulative distribution function of O3 in bins of $\mathrm{M}_{\ast}$. The primary sample of $0.6 \lesssim z \lesssim 1.0$ LEGA-C galaxies from Section~\ref{Sample} is shown by the purple curve. The mass-matched low-redshift comparison sample of $z \sim 0$ SDSS galaxies from Section~\ref{SectionThreeOne} is shown by the orange curve. The intermediate-redshift comparison sample of $z \sim 1.6$ FMOS-COSMOS galaxies from Section~\ref{FMOS} is shown by the blue curve. The high-redshift comparison sample of $z \sim 2$ KBSS galaxies from Section~\ref{KBSS} is shown by the green curve. In each of the panels, the demarcation line for $z = 0.7$ star-forming/composite galaxies and AGN from \citet{Juneau:2014} is shown in purple. Galaxies at $z = 0.7$ to the left of these regions are likely powered by star formation while galaxies to the right are more likely to contain AGN.}
    \label{fig:MEX_O3_bins_CDF}
\end{figure*}

Overall, we find that the locus of star-forming $0.6 \lesssim z \lesssim 1.0$ LEGA-C galaxies appears qualitatively much more similar to $z \sim 0$ SDSS galaxies than $z \sim 1.6$ FMOS-COSMOS or $z \sim 2$ KBSS galaxies, generally exhibiting lower O3 at fixed $\mathrm{M}_{\ast}$ than the $z \sim 1.6$ and $z \sim 2$ galaxies. However, in order to fairly compare the O3 distributions of LEGA-C and SDSS galaxies, the $\mathrm{M}_{\ast}$ distributions must be consistent with one another, particularly at low $\mathrm{M}_{\ast}$ where incompleteness in the LEGA-C sample is more of a concern. To create a mass-matched sample of SDSS galaxies, we randomly sample $N = 10$ stellar masses from a Gaussian distribution centered on the stellar mass of each galaxy within the primary sample of LEGA-C galaxies. The standard deviation of this distribution is set to be the characteristic errors on the $\mathrm{M}_{\ast}$ from LEGA-C \citep[$\simeq 0.1 - 0.2$~dex;][]{Muzzin:2013a, Muzzin:2013b}. We then pick the SDSS galaxy with the closest stellar mass to each of the randomly sampled stellar masses, without replacement. This results in a low-redshift mass-matched comparison sample of $N = 7670$ galaxies, described in Tables~\ref{tab:samples} and \ref{tab:properties}. \par

Figure~\ref{fig:histogram_mass} shows the distribution of $\mathrm{M}_{\ast}$ for galaxies within the primary sample of LEGA-C galaxies from Section~\ref{Sample} (shown in purple) and the low-redshift mass-matched comparison sample of SDSS galaxies (shown in orange). We compare the $\mathrm{M}_{\ast}$ distributions of LEGA-C and SDSS galaxies by performing Kolmogorov$-$Smirnov (KS) and Anderson$-$Darling (AD) tests on the primary sample of LEGA-C galaxies and the low-redshift mass-matched comparison sample of SDSS galaxies. These are two-sided tests for the null hypothesis that two independent samples are drawn from the same continuous distribution, with the KS test being more sensitive to the center of the distribution and the AD test being more sensitive to the tails of the distribution. The results of these tests indicate that the stellar masses of the LEGA-C and SDSS samples shown in Figure~\ref{fig:histogram_mass} are consistent with being drawn from the same parent distribution. \par

To better understand and quantify the similarities and differences between the four samples of galaxies shown in Figure~\ref{fig:MEX_O3}, we compare the cumulative distribution functions (CDF) of O3 in four bins of $\mathrm{M}_{\ast}$, the boundaries of which are illustrated by the black dotted lines in Figures~\ref{fig:MEX_O3} and \ref{fig:histogram_mass}. These bins of $\mathrm{M}_{\ast}$ were chosen to sample galaxies just below the mass-completeness limit of LEGA-C and to contain roughly the same number of galaxies in each bin. This comparison is presented in Figure~\ref{fig:MEX_O3_bins_CDF}, with the primary sample of $0.6 \lesssim z \lesssim 1.0$ LEGA-C galaxies from Section~\ref{Sample} shown by the purple curve. The low-redshift mass-matched comparison sample of $z \sim 0$ SDSS galaxies is shown by the orange curve. The intermediate-redshift comparison sample of $z \sim 1.6$ FMOS-COSMOS galaxies from Section~\ref{FMOS} is shown by the blue curve. The high-redshift comparison sample of $z \sim 2$ KBSS galaxies from Section~\ref{KBSS} is shown by the green curve. The demarcation between $z = 0.7$ star-forming/composite galaxies and AGN from \citet{Juneau:2014} is shown in purple for each of the $\mathrm{M}_{\ast}$ bins. Galaxies to the left of these shaded regions are likely powered by star formation while galaxies to the right are more likely to contain AGN. \par

% Table summarizing [O\,III]/H\beta percentiles
\begin{sidewaystable}
	\centering
	\caption{Summary of O3 percentiles in bins of stellar mass.}
	\label{tab:O3_statistics}
	\begin{tabular}{ccccccccccccc}
		\hline
		\hline
		& \multicolumn{3}{c}{SDSS} & \multicolumn{3}{c}{LEGA-C} & \multicolumn{3}{c}{FMOS-COSMOS} & \multicolumn{3}{c}{KBSS} \\
		\hline
		Mass Bin & 16th & 50th & 84th & 16th & 50th & 84th & 16th & 50th & 84th & 16th & 50th & 84th \\
		\hline
		$10.00 < \mathrm{log}(\mathrm{M_{\ast}/M_{\odot}}) < 10.25$ & $-$0.607 & $-$0.398 & $-$0.126 & $-$0.459 & $-$0.260 & $+$0.010 & $-$0.029 & $+$0.222 & $+$0.464 & $+$0.300 & $+$0.503 & $+$0.654 \\
		$10.25 < \mathrm{log}(\mathrm{M_{\ast}/M_{\odot}}) < 10.50$ & $-$0.614 & $-$0.376 & $+$0.070 & $-$0.498 & $-$0.287 & $+$0.061 & $-$0.150 & $+$0.093 & $+$0.352 & $+$0.196 & $+$0.432 & $+$0.647 \\
		$10.50 < \mathrm{log}(\mathrm{M_{\ast}/M_{\odot}}) < 10.75$ & $-$0.589 & $-$0.274 & $+$0.237 & $-$0.537 & $-$0.242 & $+$0.112 & $-$0.176 & $+$0.027 & $+$0.377 & $+$0.170 & $+$0.350 & $+$0.608 \\
		$10.75 < \mathrm{log}(\mathrm{M_{\ast}/M_{\odot}}) < 11.00$ & $-$0.519 & $-$0.077 & $+$0.374 & $-$0.484 & $-$0.126 & $+$0.468 & $-$0.136 & $+$0.106 & $+$0.369 & $+$0.004 & $+$0.285 & $+$0.877 \\
		\hline
	\end{tabular}
\end{sidewaystable}

In Figure~\ref{fig:MEX_O3_bins_CDF}, we now see \emph{quantitatively} that the LEGA-C galaxies are indeed more similar to $z \sim 0$ SDSS galaxies than $z \sim 1.6$ FMOS-COSMOS or $z \sim 2$ KBSS galaxies at all $\mathrm{M}_{\ast}$. The 16th, 50th, and 84th percentiles in O3 for all samples are provided in Table \ref{tab:O3_statistics}. Looking at the median O3 values of the samples, the LEGA-C galaxies have significantly lower values of O3 at all $\mathrm{M}_{\ast}$ when compared to FMOS-COSMOS and KBSS galaxies (up to 0.8~dex lower in the lowest mass bin and up to 0.4~dex lower in the highest mass bin). At high $\mathrm{M}_{\ast}$, the LEGA-C and SDSS samples are virtually identical, with LEGA-C galaxies offset toward higher values of O3 by $\lesssim 0.1$~dex. Differences increase toward lower $\mathrm{M}_{\ast}$, with a $\gtrsim 0.1$~dex offset toward higher values of O3 in the lowest mass bin. We see similar results when looking at the mean O3 values. \par

% Table summarizing [O\,III]/H\beta KS and AD tests
\begin{table}
	\centering
	\caption{Summary of O3 Kolmogorov-Smirnov and Anderson-Darling test results comparing galaxies from SDSS and LEGA-C in bins of stellar mass.}
	\label{tab:O3_tests}
	\setlength{\tabcolsep}{15pt}
	\begin{tabular}{ccc}
	    \hline
		\hline
		Mass Bin & $p_{\mathrm{KS}}$ & $p_{\mathrm{AD}}$ \\
		\hline
		$10.00 < \mathrm{log}(\mathrm{M_{\ast}/M_{\odot}}) < 10.25$ & 0.000 & 0.001 \\
		$10.25 < \mathrm{log}(\mathrm{M_{\ast}/M_{\odot}}) < 10.50$ & 0.000 & 0.001 \\
		$10.50 < \mathrm{log}(\mathrm{M_{\ast}/M_{\odot}}) < 10.75$ & 0.192 & 0.131 \\
		$10.75 < \mathrm{log}(\mathrm{M_{\ast}/M_{\odot}}) < 11.00$ & 0.906 & 0.250 \\
		\hline
	\end{tabular}
\end{table}

We also perform KS and AD tests to better compare the detailed O3 distributions of galaxies from SDSS and LEGA-C. The results of this analysis are shown in Table \ref{tab:O3_tests}. We consider the samples to be significantly different when $p \leq 0.003$\footnote{The values for AD test $p$-values are floored at 0.001 and capped at 0.250, based on the \texttt{scipy} routine we used to perform this test.}, which approximately corresponds to a significance of $3\sigma$. The $p$-values reported in Table \ref{tab:O3_tests} confirm that the mass-matched SDSS and LEGA-C samples are consistent with being drawn from the same parent (O3) population at high $\mathrm{M}_{\ast}$, but exhibit statistically-significant differences in the two lower $\mathrm{M}_{\ast}$ bins. \par

The absence of statistical differences in O3 between LEGA-C and SDSS galaxies at $10.50 < \mathrm{log}(\mathrm{M_{\ast}/M_{\odot}}) < 11.00$ indicates that the ISM conditions of $z \sim 1$ galaxies with the highest $\mathrm{M}_{\ast}$ are already similar to those found in present-day galaxies, despite a lookback time of $6-8$~Gyr. The slight but statistically significant offsets toward higher O3 at $10.00 < \mathrm{log}(\mathrm{M_{\ast}/M_{\odot}}) < 10.50$ occur near the mass-completeness limit of LEGA-C. However, if real, these differences suggest that the similarity of $z \sim 1$ galaxies to $z \sim 0$ galaxies does not extend across the entire population. Indeed, at lower $\mathrm{M}_{\ast}$ than we explore here ($8.00 < \mathrm{log}(\mathrm{M_{\ast}/M_{\odot}}) < 9.00$), $z \sim 1$ DEEP2 galaxies also appear dissimilar to the $z \sim 0$ SDSS galaxies, with offsets toward higher values of O3 \citep[][]{Ly:2015}, qualitatively consistent with the offset we see at low $\mathrm{M}_{\ast}$ in the MEx for LEGA-C in Figure~\ref{fig:MEX_O3}. These results do not change substantially if we relax the S/N threshold on the emission line measurements from LEGA-C. \par

\subsection{The Mass-O32 Diagram}
\label{SectionThreeTwo}

In addition to the MEx, it is useful to consider other emission line diagnostic diagrams that are sensitive to nebular conditions. O32 (as defined in Table~\ref{tab:definitions}) is often used as a proxy for ionization \citep{Penston:1990} and can be used to calculate the ionization parameter given an assumed ionizing spectrum \citep{Maiolino:2008}. The ionization parameter is defined here as the dimensionless ratio of local hydrogen-ionizing photon density to local hydrogen densities. Thus, we can attribute differences in O32 to differences in (1) ISM densities or (2) ionizing photon densities (which could be caused by, e.g., spatial clustering of star formation). O32 is seen to evolve significantly with redshift, with higher values of O32 at fixed $\mathrm{M}_{\ast}$, potentially signaling important shifts in these physical properties \citep[e.g.,][]{Sanders:2016}. \par

In contrast to O3, O32 requires nebular reddening corrections as a result of the widely separated wavelengths of the lines used ([O\,II]$\mathrm{\lambda\lambda 3727,3729}$ and [O\,III]$\mathrm{\lambda\lambda 4960,5008}$). We use the Balmer decrement H$\gamma$/H$\beta$ to apply these corrections to all of the samples, as these are the strongest Balmer lines available from the VIMOS spectra, assuming an intrinsic value of H$\gamma$/H$\beta$ = 0.47 \citep[][]{Osterbrock:2006}. Typical values of extinction are $0.0 < E(B - V)/\mathrm{mag} < 0.4$. \par

\begin{figure}
    \centering
	\includegraphics[width=0.95\linewidth]{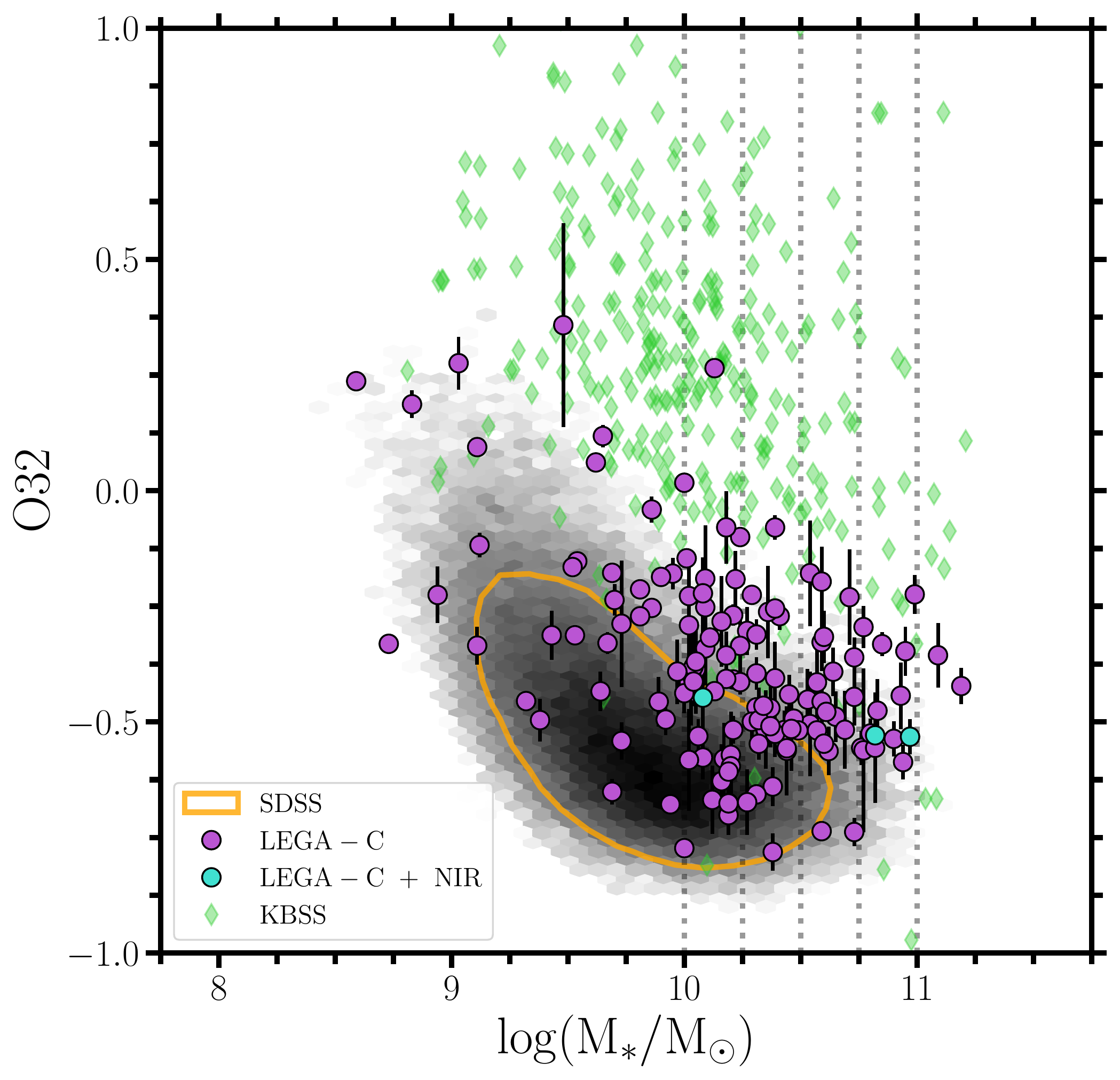}
    \caption{The $\mathrm{M}_{\ast}$-O32 diagram. The secondary sample of $0.6 \lesssim z \lesssim 1.0$ LEGA-C galaxies from Section~\ref{Sample} is shown with the purple circles. The subsample of LEGA-C galaxies from Section~\ref{Subsample} with observed-NIR spectra is shown with the cyan circles. The low-redshift comparison sample of $z \sim 0$ SDSS galaxies from Section~\ref{SDSS} is shown in grayscale with an orange contour, where the contour encloses roughly 90\% of the low-redshift sample. The high-redshift comparison sample of $z \sim 2$ KBSS galaxies from Section~\ref{KBSS} is shown with the green diamonds. The black dotted vertical lines indicate the $\mathrm{M}_{\ast}$ bins that are used throughout. While the $0.6 \lesssim z \lesssim 1.0$ LEGA-C galaxies appear more similar to the $z \sim 0$ SDSS galaxies than the $z \sim 2$ KBSS galaxies in this parameter space as well, they are still offset toward higher O3 by $\simeq 0.1 - 0.2$~dex in all mass bins.}
    \label{fig:O32_mass}
\end{figure}

In Figure~\ref{fig:O32_mass} we present the $\mathrm{M}_{\ast}$-O32 diagram. The secondary sample of $0.6 \lesssim z \lesssim 1.0$ LEGA-C galaxies from Section~\ref{Sample} is shown with purple circles. The secondary sample is used here since O32 additionally requires $\mathrm{S/N} > 3$ measurements of [O\,II]$\mathrm{\lambda\lambda 3727,3729}$ and [O\,III]$\mathrm{\lambda 4960}$ after continuum subtraction. The subsample of LEGA-C galaxies from Section~\ref{Subsample} with observed-NIR spectra is shown with the cyan circles. \par

The low-redshift comparison sample of $z \sim 0$ SDSS galaxies from Section~\ref{SDSS} is shown in grayscale with an orange contour, where the contour encloses roughly 90\% of the low-redshift sample. The high-redshift comparison sample of $z \sim 2$ KBSS galaxies from Section~\ref{KBSS} is shown with the green diamonds. Because FMOS-COSMOS did not routinely observe [O\,II]$\mathrm{\lambda\lambda 3727,3729}$ for their $z \sim 1.6$ galaxies, we do not include that sample in this comparison. \par

We exclude SDSS galaxies that fall above the empirical curve from \citet{Kauffmann:2003} and KBSS galaxies that were identified as AGN as described in \citet{Strom:2017}. LEGA-C galaxies that are identified as AGN as described in Section~\ref{SectionFourTwo} are also excluded. This includes LEGA-C galaxies that are to the right of the demarcation line from \citet{Juneau:2014} or have significant hard X-ray emission ${\mathrm{log}( \mathrm{L_{\,2-10\,\mathrm{keV}}} / [\mathrm{erg/s}] )} > 42.0$. \par

In this parameter space, as in the MEx, $0.6 \lesssim z \lesssim 1.0$ LEGA-C galaxies appear much more similar to $z \sim 0$ SDSS galaxies than $z \sim 2$ KBSS galaxies. Looking at the median O32 values of the samples in the same bins as shown in Figure~\ref{fig:histogram_mass} (also shown here using vertical dotted lines), the LEGA-C galaxies have significantly lower values of O32 when compared to KBSS galaxies: up to 0.7~dex lower in the lowest mass bin and up to 0.4~dex lower in the highest mass bin. The LEGA-C galaxies have somewhat higher values of O32 when compared to SDSS galaxies: up to 0.2~dex higher in the lowest mass bin and up to 0.1~dex higher in the highest mass bin. We see similar results when looking at the mean O32 values. These results might suggest marginally higher nebular ionization in $0.6 \lesssim z \lesssim 1.0$ LEGA-C galaxies when compared to $z \sim 0$ SDSS galaxies, particularly at low $\mathrm{M}_{\ast}$. \par

% O32-mass Diagram

\subsection{The Mass-R23 and Mass-Metallicity Diagrams}
\label{SectionThreeThree}

The R23 index (defined in Table~\ref{tab:definitions}) is often used to estimate gas-phase oxygen abundance \citep{Pagel:1979}, given an additional parameter to break the double-valued degeneracy of R23 with O/H \citep{Maiolino:2008}. R23 can also be used as a probe of excitation as it compares the emission of collisionally-excited metal lines with a recombination hydrogen line, similar to O3. However, R23 is less sensitive to differences in ionization than O3 because it includes both [O\,II] and [O\,III]. Thus, we can attribute differences in the $\mathrm{M}_{\ast}$-R23 diagram to differences in (1) gas-phase metallicity and/or (2) photoionization heating from massive stars. \par 

Like O32, R23 requires nebular reddening corrections as a result of the relative wavelengths of the lines used ([O\,II]$\mathrm{\lambda\lambda 3727,3729}$, H$\beta$, and [O\,III]$\mathrm{\lambda\lambda 4960,5008}$). As described in Section~\ref{SectionThreeTwo}, we use the Balmer decrement H$\gamma$/H$\beta$ to apply these corrections to all of the samples. \par

\begin{figure}
    \centering
	\includegraphics[width=0.95\linewidth]{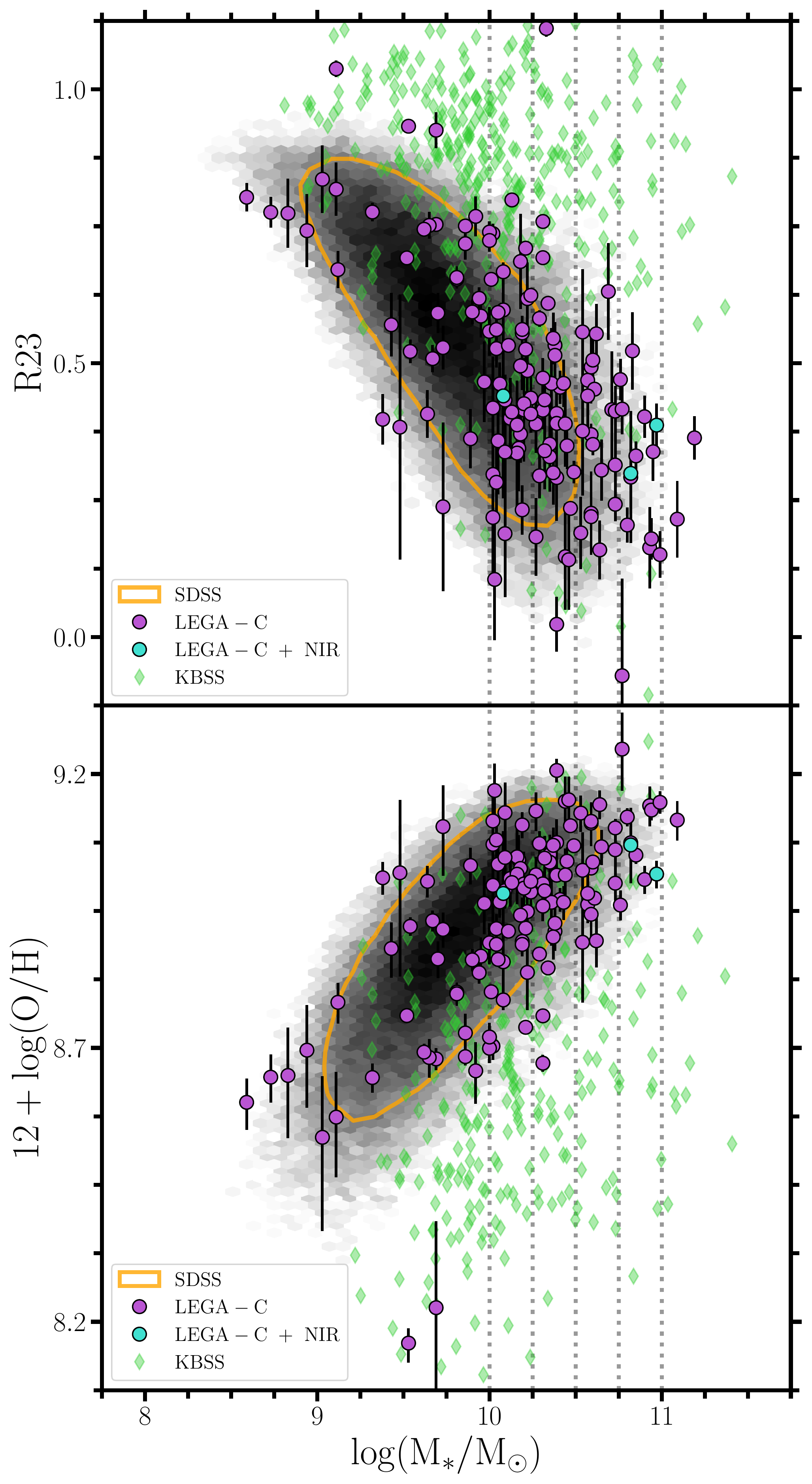}
    \caption{Shown in the top panel, the $\mathrm{M}_{\ast}$-R23 diagram. Shown in the bottom panel, the $\mathrm{M}_{\ast}$-O/H diagram. The secondary sample of $0.6 \lesssim z \lesssim 1.0$ LEGA-C galaxies from Section~\ref{Sample} is shown with the purple circles. The subsample of LEGA-C galaxies from Section~\ref{Subsample} with observed-NIR spectra is shown with the cyan circles. The low-redshift comparison sample of $z \sim 0$ SDSS galaxies from Section~\ref{SDSS} is shown in grayscale with an orange contour, where the contour encloses roughly 90\% of the low-redshift sample. The high-redshift comparison sample of $z \sim 2$ KBSS galaxies from Section~\ref{KBSS} is shown with the green diamonds. The black dotted vertical lines indicate the $\mathrm{M}_{\ast}$ bins that are used throughout. We see that the $0.6 \lesssim z \lesssim 1.0$ LEGA-C galaxies are substantially more enriched than the $z \sim 2$ KBSS galaxies, with virtually identical ranges in $\mathrm{M}_{\ast}$-O/H as the $z \sim 0$ SDSS galaxies.}
    \label{fig:Z_R23}
\end{figure}

In the top panel of Figure~\ref{fig:Z_R23}, we present the $\mathrm{M}_{\ast}$-R23 diagram. The secondary sample of $0.6 \lesssim z \lesssim 1.0$ LEGA-C galaxies from Section~\ref{Sample} is shown with purple circles. The secondary sample is used here since O32 additionally requires $\mathrm{S/N} > 3$ measurements of [O\,II]$\mathrm{\lambda\lambda 3727,3729}$ and [O\,III]$\mathrm{\lambda 4960}$ after continuum subtraction. The subsample of LEGA-C galaxies from Section~\ref{Subsample} with observed-NIR spectra is shown with the cyan circles. \par

The low-redshift comparison sample of $z \sim 0$ SDSS galaxies from Section~\ref{SDSS} is shown in grayscale with an orange contour, where the contour encloses roughly 90\% of the low-redshift sample. The high-redshift comparison sample of $z \sim 2$ KBSS galaxies from Section~\ref{KBSS} is shown with the green diamonds. Like the $\mathrm{M}_{\ast}$-O32 diagram, we exclude galaxies that were identified as AGN from all samples. \par

In the bottom panel of Figure~\ref{fig:Z_R23}, we present the $\mathrm{M}_{\ast}$-O/H diagram. To calculate gas-phase O/H, we use the R23 conversion from \citet{Maiolino:2008}. We assume the higher value of gas-phase oxygen abundance in order to break the degeneracy of R23 with O/H since the LEGA-C galaxies are relatively high $\mathrm{M}_{\ast}$, with mass completeness at log($\mathrm{M_{\ast}/M_{\odot}}$) $\gtrsim 10.3$. The symbols in this panel are the same as in the top panel. \par

In this parameter space, as in the MEx and $\mathrm{M}_{\ast}$-O32 diagram, $0.6 \lesssim z \lesssim 1.0$ LEGA-C galaxies appear more similar to $z \sim 0$ SDSS galaxies than $z \sim 2$ KBSS galaxies. The LEGA-C and SDSS samples are virtually identical, with $\lesssim 0.1$~dex offsets toward lower values of O/H in all $\mathrm{M}_{\ast}$ bins. We see similar results when looking at the mean R23 values. These results imply similar gas-phase enrichment for both $0.6 \lesssim z \lesssim 1.0$ LEGA-C galaxies and $z \sim 0$ SDSS galaxies. \par

% R23-mass and Z-mass Diagram

%%%%%
\section{Discussion}
\label{SectionFour}

The first studies of large samples of high-redshift ($z \gtrsim 2$) galaxies showed that they exhibit large differences in O3 at fixed $\mathrm{M}_{\ast}$ when compared to low-redshift samples of galaxies, having much higher values at all stellar masses. Previous work has found that this offset can be explained by differences in gas-phase oxygen abundance and the characteristic shape of the ionizing radiation field of massive stars, which is largely determined by their iron enrichment \citep[e.g.,][]{Steidel:2016, Strom:2018, Shapley:2019, Topping:2020, Runco:2021}. Still, exactly how and when these differences might arise in the galaxy population remains uncertain. \par

We can learn more about the changes in galaxy chemistry and their connection to galaxy assembly histories by studying samples at intermediate redshift, when these changes would have occurred. \citet{Kashino:2019} showed that most $z \sim 1.6$ galaxies also have elevated O3 at fixed $\mathrm{M}_{\ast}$ compared to $z \sim 0$ galaxies, but that these differences were less pronounced at higher $\mathrm{M}_{\ast}$ ($\log(\mathrm{M}_{\ast}/\mathrm{M}_{\odot}) > 10.6$). \par

In Section~\ref{SectionThree}, we showed that $0.6 \lesssim z \lesssim 1.0$ galaxies with $\log(\mathrm{M}_{\ast}/\mathrm{M}_{\odot}) > 10.5$ appear much more similar to $z \sim 0$ galaxies than $z \sim 1.6$ or $z \sim 2$ galaxies, not only in the MEx (Section~\ref{SectionThreeOne}), but also using other nebular diagnostics (Sections~\ref{SectionThreeTwo} and \ref{SectionThreeThree}). This suggests that even at a lookback time of $6-8$~Gyr many galaxies had similar nebular properties to present-day galaxies, and that much of this change occurred in the $3-4$~Gyr between $z = 1.6$ and $z = 0.7$. There is some evidence that the transition is not complete at lower $\mathrm{M}_{\ast}$, although this is near the mass-completeness limit for LEGA-C. \par

Below, we consider several possible explanations (both observational and physical) for the distribution of O3 in the LEGA-C sample: (1) selection effects, (2) contributions from AGN, and (3) changes in physical conditions. Additionally, we discuss results from analysis of the full rest-optical spectra for a subsample of galaxies from LEGA-C. \par

\subsection{Selection Effects}

\citet{Juneau:2014} demonstrated the importance of accounting for selection effects caused by emission line detection limits when investigating the redshift-dependent offset of galaxies on emission-line diagnostic diagrams such as the MEx. Specifically, an overall emission line luminosity limit of ${\mathrm{log}( \mathrm{L_{line}} / [\mathrm{erg/s}] )} > 41.0$ applied to $z \sim 0$ SDSS galaxies results in a sample that exhibits the same offset toward higher O3 as seen in $z \sim 2$ KBSS galaxies. This suggests that $z \sim 0$ SDSS galaxies with the largest line luminosities could be considered low-redshift analogues, if typical high-$z$ galaxies are more luminous than typical low-$z$ galaxies at fixed $\mathrm{M}_{\ast}$ \citep[][]{Cowie:2016, Bian:2016}. At the same time, selection effects alone could create an apparent enhancement in O3 at fixed $\mathrm{M}_{\ast}$, if $z \gg 0$ samples are biased toward galaxies with the largest intrinsic line luminosities. \par

\begin{figure}
    \centering
	\includegraphics[width=1.0\linewidth]{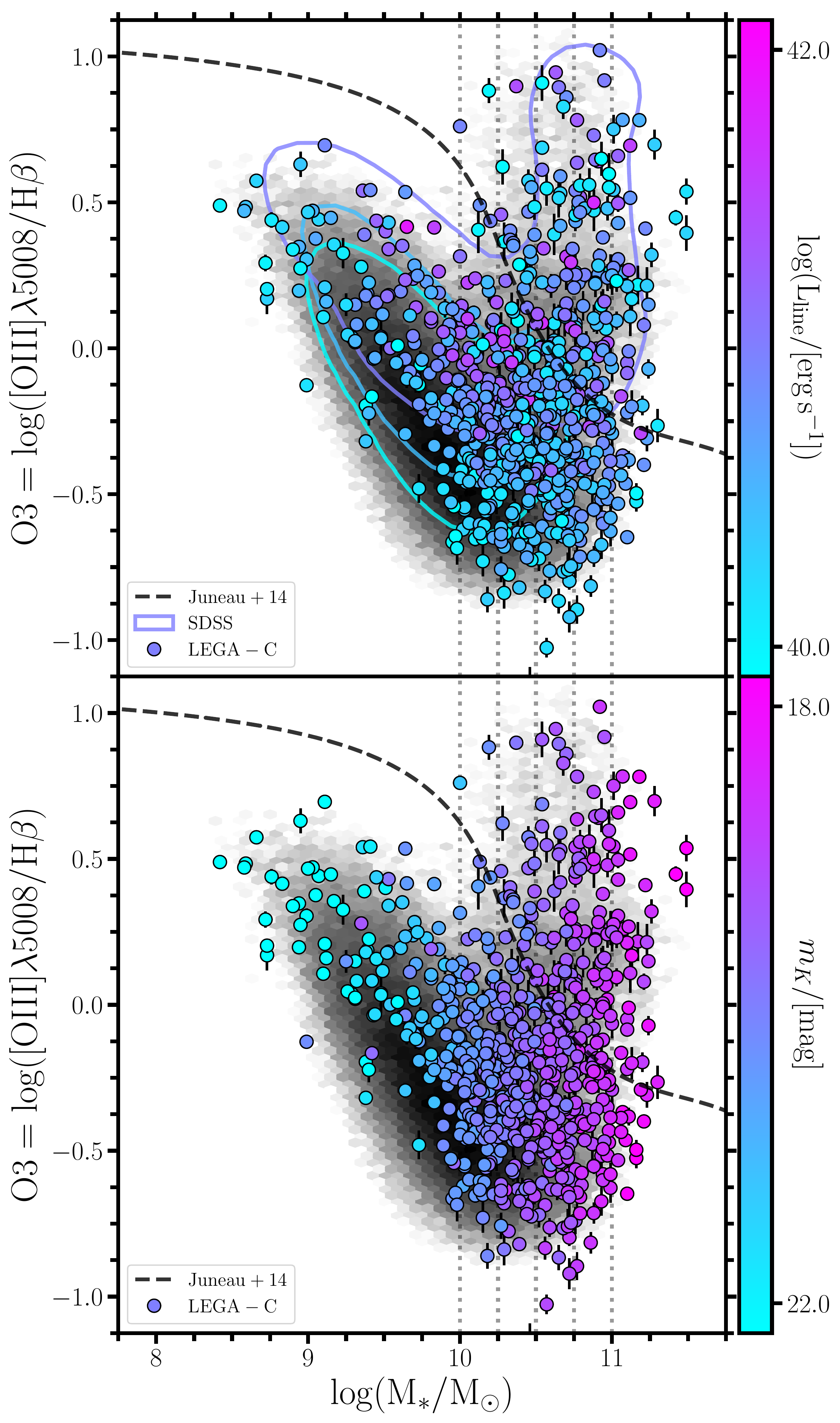}
    \caption{Shown in the top panel, the MEx diagram with the primary sample of $0.6 \lesssim z \lesssim 1.0$ LEGA-C galaxies from Section~\ref{Sample} color-coded based on limiting line luminosity. The comparison sample of $z \sim 0$ SDSS galaxies from Section~\ref{SDSS} is shown in grayscale with contours that correspond to increasing individual line luminosity thresholds applied to H$\beta$ and [O\,III]$\mathrm{\lambda 5008}$, where each contour encloses roughly 90\% of the low-redshift sample greater than or equal to that luminosity. Shown in the bottom panel, the MEx diagram with the primary sample of $0.6 \lesssim z \lesssim 1.0$ LEGA-C galaxies color-coded based on their $K$-band magnitude. The black dotted vertical lines in both panels indicate the $\mathrm{M}_{\ast}$ bins that are used in the analysis throughout the text. We see that the LEGA-C galaxies in our sample exhibit the same trend as SDSS, with higher emission line luminosities corresponding to higher values of O3. However, we also see that the $K$-band magnitude selection of the parent LEGA-C sample is not significantly biasing the distribution at low O3, where we see some differences relative to SDSS at $10.00 < \mathrm{log}(\rm M_{\ast}/\rm M_{\odot}) < 10.50$.}
    \label{fig:MEX_O3_luminosity}
\end{figure}

Looking at trends in emission line luminosities in the MEx can help us assess whether selection effects and observational biases have a significant impact on the distribution of LEGA-C galaxies. We would expect LEGA-C galaxies with the highest O3 values to exhibit the largest line luminosity if the intermediate-$z$ galaxies follow the same trends as $z \sim 0$ galaxies. In the top panel of Figure~\ref{fig:MEX_O3_luminosity}, we present the MEx diagram again, now with the primary sample of $0.6 \lesssim z \lesssim 1.0$ LEGA-C galaxies from Section~\ref{Sample} color-coded based on their limiting line luminosity. The low-redshift comparison sample of $z \sim 0$ SDSS galaxies from Section~\ref{SDSS} is shown in grayscale with contours that correspond to increasing line luminosity thresholds applied to both H$\beta$ and [O\,III]$\mathrm{\lambda 5008}$, where each contour encloses roughly 90\% of the low-redshift sample satisfying that selection. \par 

% Mass-Excitation Diagram with line luminosity gradient

We see that $z \sim 0$ SDSS galaxies with larger emission line luminosities tend to have higher values of O3, which is the same trend discussed in \citet{Juneau:2014}. We additionally find that $0.6 \lesssim z \lesssim 1.0$ LEGA-C galaxies exhibit the same trend in the three lowest mass bins ($10.00 < \mathrm{log}(\rm M_{\ast}/\rm M_{\odot}) < 10.75$) but no trend in the highest mass bin ($10.75 < \mathrm{log}(\rm M_{\ast}/\rm M_{\odot}) < 11.00$) according to a Spearman rank correlation test. This is contrary to observations at $z \sim 2$, where the degree of offset from $z \sim 0$ galaxies in diagnostics diagrams like the BPT diagram is not correlated with, e.g., Balmer line luminosity \citep{Strom:2017}. Since there is no observed trend in the highest mass bin, observational biases likely do not have a significant impact on the locations of intermediate-$z$ galaxies on the MEx at $10.75 < \mathrm{log}(\rm M_{\ast}/\rm M_{\odot}) < 11.00$. However, the correlation between emission line luminosity and O3 at low $\mathrm{M}_{\ast}$ in the LEGA-C sample could imply that selection effects and/or observational biases may play a role in the observed distribution of O3. Specifically, the differences seen in Figure~\ref{fig:MEX_O3_bins_CDF} could be explained if LEGA-C is preferentially biased against galaxies with lower line luminosities. Conversely, we might expect some evolution in typical line luminosity, given the overall increase in SFR at fixed $\mathrm{M}_{\ast}$ with increasing redshift \citep[][]{Juneau:2014}, which could lead to differences in O3. \par

To disentangle the impact of selection effects and redshift evolution---which is ultimately what we are aiming to characterize---we check to see if O3 is correlated with $K$-band magnitude, which was one of the criteria originally used to select galaxies in LEGA-C, along with photometric redshift. In the bottom panel of Figure~\ref{fig:MEX_O3_luminosity}, we present the MEx diagram again, now with the primary sample of $0.6 \lesssim z \lesssim 1.0$ LEGA-C galaxies from Section~\ref{Sample} color-coded based on their $K$-band magnitude. We see that brighter LEGA-C galaxies tend to have higher $\mathrm{M}_{\ast}$, as expected \citep[][]{LEGAC_DR3}. At fixed $\mathrm{M}_{\ast}$, there is no significant correlation between $K$-band magnitude and O3 in the lowest mass bin ($10.00 < \mathrm{log}(\rm M_{\ast}/\rm M_{\odot}) < 10.25$), but a Spearman test reveals a weak correlation in the three highest mass bins ($10.25 < \mathrm{log}(\rm M_{\ast}/\rm M_{\odot}) < 11.00$): at these masses, fainter galaxies generally have \emph{higher} O3, suggesting that the parent sample selection is not biasing the \emph{lower} envelope of the $z \sim 1$ locus in the MEx diagram. Further, as we noted in Section~\ref{SectionThreeOne}, the emission line S/N threshold we adopt does not have a significant impact on the observed distribution of O3, and so we are effectively capturing the behavior of the observed LEGA-C sample. Together, these results suggest that the slight offset toward higher O3 at lower $\mathrm{M}_{\ast}$ is not the result of selection effects or observational biases. \par

\subsection{Contributions from AGN}
\label{SectionFourTwo}

Contributions from AGN can also have a significant impact on the location of galaxies in the MEx. The harder ionizing spectra of AGN result in high values of O3 and are responsible for the location of galaxies in the upper-right portion of the MEx diagram. This means that the offset toward higher O3 observed for LEGA-C at low $\mathrm{M}_{\ast}$ could be explained by a comparatively larger number of composite galaxies or AGN present in the LEGA-C sample at those masses. Thus, it is important to understand the likelihood of contributions from AGN given the observed similarities and differences between LEGA-C and SDSS. \par

The MEx demarcation line between $z = 0.7$ star-forming/composite galaxies and AGN from \citet{Juneau:2014} is represented by the purple shaded regions in Figure~\ref{fig:MEX_O3_bins_CDF}. As this curve was designed to identify even heavily-absorbed AGN and calibrated using both $z \sim 0$ galaxies from SDSS and $z \sim 1$ galaxies from TKRS and DEEP2, we believe significant contributions from AGN in the two lowest mass bins ($10.00 < \mathrm{log}(\rm M_{\ast}/\rm M_{\odot}) < 10.50$) are minor. In these mass bins, the majority of LEGA-C galaxies are to the left of the demarcation line and are likely powered by star formation. \par

% Mass-Excitation Diagram with X-ray data
\begin{figure}
    \centering
	\includegraphics[width=0.95\linewidth]{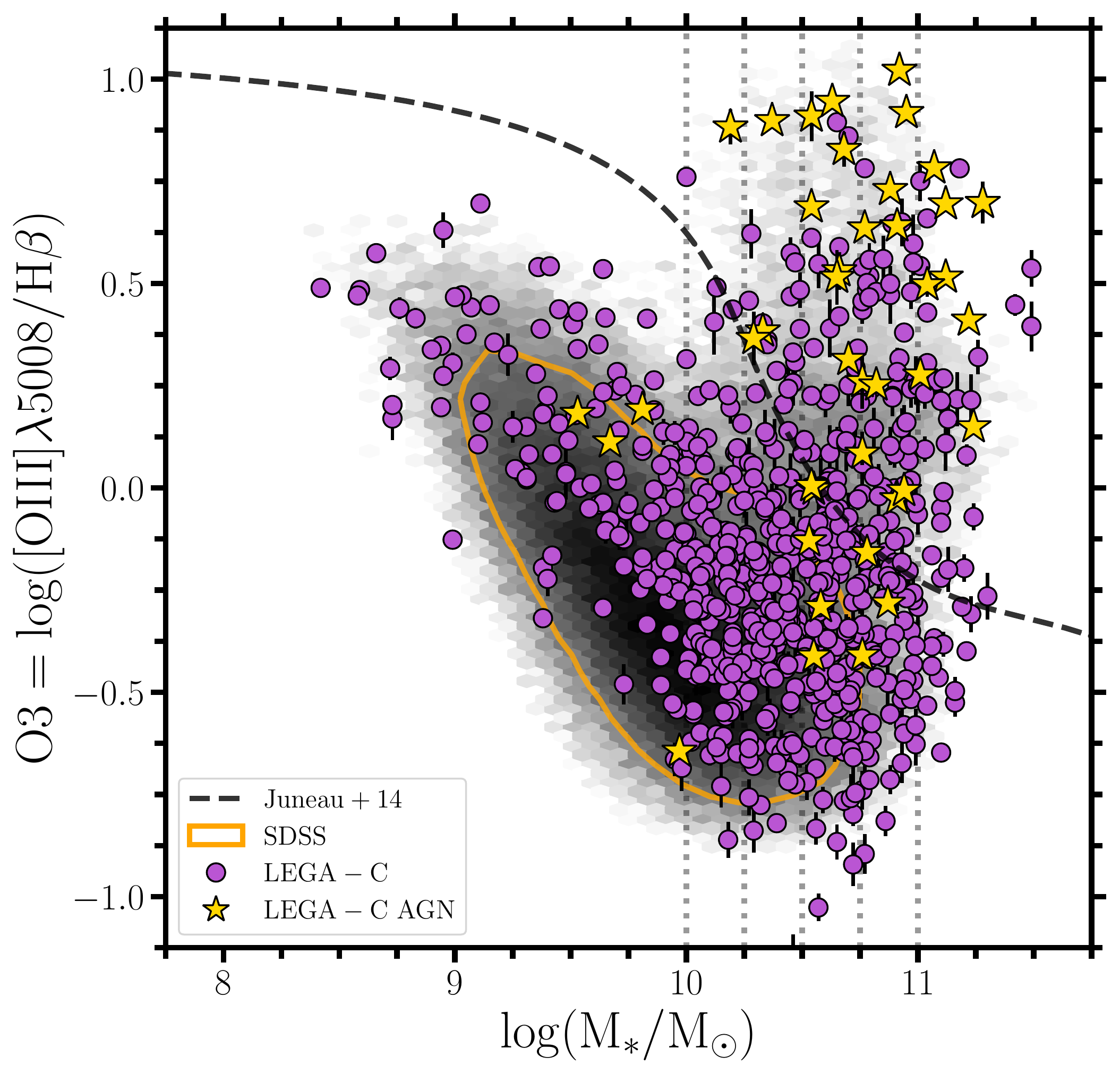}
    \caption{The MEx. The primary sample of $0.6 \lesssim z \lesssim 1.0$ LEGA-C galaxies from Section~\ref{Sample} is shown with the purple circles. Galaxies within the primary sample that are identified as AGN by their X-ray luminosities are marked with yellow stars. The low-redshift comparison sample is shown as in Figure~\ref{fig:MEX_O3}. The division between $z = 0.7$ star-forming/composite galaxies and AGN from \citet{Juneau:2014} is given by the black dashed line. The black dotted vertical lines indicate the $\mathrm{M}_{\ast}$ bins that are used throughout.}
    \label{fig:MEX_O3_xray}
\end{figure}

Another way to identify potential AGN is by using hard X-ray emission, with ${\mathrm{log}( \mathrm{L_{\,2-10\,\mathrm{keV}}} / [\mathrm{erg/s}] )} > 42.0$ serving as the nominal luminosity threshold used at $z < 1$ \citep[][]{Juneau:2011}. In Figure~\ref{fig:MEX_O3_xray} we present the MEx diagram with the primary sample of $0.6 \lesssim z \lesssim 1.0$ LEGA-C galaxies from Section~\ref{Sample} marked with yellow stars if they are identified as potential AGN from their \textit{Chandra} hard X-ray emission \citep[][]{Civano:2016, Marchesi:2016}. The majority of $0.6 \lesssim z \lesssim 1.0$ LEGA-C galaxies identified as X-ray AGN (30/42 galaxies) are also identified as AGN by the demarcation line from \citet{Juneau:2014}, which supports using the curve from \citet{Juneau:2014} for intermediate-$z$ samples. In Sections~\ref{SectionThreeTwo} and \ref{SectionThreeThree}, we removed all LEGA-C galaxies that are identified as AGN, using either the \citet{Juneau:2014} classification or on the basis of their X-ray luminosities. \par

Although lower luminosity AGN could still be present at low $\mathrm{M}_{\ast}$, there is reason to believe that they are not responsible for the range of O3 at these masses, as low-luminosity AGN do not dominate the observed strong optical emission lines in low-mass star-forming galaxies \citep[][]{Trump:2015}. The challenges associated with identifying AGN as a potential ionizing source demonstrate the need for full access to the rest-optical spectrum of intermediate-$z$ galaxies. Although low-mass AGN may still have line ratios consistent with the star forming sequence in, e.g., the BPT diagram \citep[][]{Baldassare:2018}, using multiple methods will nonetheless help quantify the contributions from AGN in intermediate-$z$ samples. We revisit this issue in Section~\ref{Future}, where we include observed-NIR spectra, allowing for complete coverage of the rest-optical for a subsample of the LEGA-C galaxies. \par

\subsection{Physical Conditions}
\label{SectionFourThree}

Differences in galaxies' nebular spectra can also be caused by differences in physical conditions, such as ionization parameter and/or gas-phase oxygen abundance. Having investigated the impact of selection effects and contributions from AGN on the location of intermediate-$z$ LEGA-C galaxies in the MEx diagram, we now consider the role of ionization and enrichment, which we previously discussed in Sections~\ref{SectionThreeTwo} and \ref{SectionThreeThree}. \par

To understand the impact of changing physical conditions, we can borrow from the theoretical approach of \citet{Kewley:2013b}, who predicted the location of $z > 0$ galaxies on the N2-BPT diagram. Their framework invokes two sequences to explain the appearance of the galaxy locus in the N2-BPT diagram: an abundance sequence corresponding to the star-forming branch, and a mixing sequence where the AGN contribution becomes significant. \citet{Kewley:2013b} use the photoionization code MAPPINGS IV \citep{Dopita:2013} to define the abundance sequence, and MAPPINGS III \citep{Groves:2004} included the effects of AGN, demonstrating that an increasing AGN contribution forms the mixing sequence. This framework also describes the behavior of $z \sim 0$ galaxies in the MEx, because it quantifies the effect on both N2 and O3. \par

In the \citet{Kewley:2013b} model, raising the ionization parameter at lower metallicities results in higher values of O3. At fixed $\mathrm{M_{\ast}}$, we find that LEGA-C galaxies exhibit a $\simeq 0.1 - 0.2$~dex offset toward higher O32 at fixed $\mathrm{M}_{\ast}$ when compared to SDSS galaxies but are not offset in O/H ($\lesssim 0.1$~dex). We would expect this combination of parameters to result in an offset toward higher O3 at fixed $\mathrm{M}_{\ast}$, however, we only see statistical evidence of any potential differences at the lowest $\mathrm{M}_{\ast}$ probed by LEGA-C. This may suggest that the relationship between ionization and enrichment in galaxies at these redshifts differs from the models, but confirming or rejecting this requires better constraints on both ionization parameter and metallicity for a large sample. Detailed photoionization modeling of the sample considered here is the subject of forthcoming work. \par

Since we only see statistical evidence of differences in O3 at the lowest M$_{\ast}$ probed by LEGA-C, our results suggest that low-mass intermediate-redshift galaxies are less ``evolved'' than their high-mass counterparts. This is consistent with the results from \citet{Kashino:2019}, who also reported more similarities between $z \sim 1.6$ and $z \sim 0$ galaxies at high $\mathrm{M}_{\ast}$. Together, these findings support a version of the ``down-sizing'' picture proposed by \citet[][]{Cowie:1996}, who postulated that the stars in more massive galaxies tend to have formed earlier and over a shorter period of time compared to the stars in less massive galaxies. Additionally, they found that the mass assembly process in the highest density regions was largely complete by $z \sim 1$, which agrees with our analysis of the nebular spectra of LEGA-C galaxies.\par 

\subsection{Full Rest-Optical}
\label{Future}

Given our results using the partial rest-optical spectra of a large sample of $0.6 < z < 1.0$ LEGA-C galaxies (introduced in Section~\ref{Sample}), it is important for us to understand how these results compare to results from the full rest-optical spectra for a smaller subsample of $0.6 < z < 1.0$ LEGA-C galaxies that have NIR observations (introduced in Section~\ref{Subsample}). Throughout, we have shown this subsample alongside the larger sample to see this comparison. Here, we further investigate these galaxies. \par

\subsubsection{The Mass-N2O2 Diagram}
\label{SectionFourFourOne}

N2O2 (as defined in Table~\ref{tab:definitions}) can be used as a proxy for nitrogen-to-oxygen ratio for galaxies at a variety of redshifts \citep[N/O;][]{Perez-Montero:2009, Pilyugin:2012, Strom:2018} since [N\,II] and [O\,II] have similar ionization potentials ($14.5\, \mathrm{eV}$ for [N\,II] and $13.6\, \mathrm{eV}$ for [O\,II]). Thus, we can attribute differences in values of N2O2 to differences in N/O. N2O2 is a complementary probe of nebular enrichment that is worth pursuing in addition to R23, because N2O2 can be used more reliably at all redshifts (particularly at $z \gg 0$). Similar ionization potentials make N2O2 relatively insensitive to changes in the ionizing spectrum, either with differences in the ionization parameter or differences in the shape of the ionizing radiation. However, N2O2 does require nebular extinction corrections; again, the Balmer decrement H$\gamma$/H$\beta$ is used to apply these corrections to all of the samples. \par

\begin{figure}
    \centering
	\includegraphics[width=0.95\linewidth]{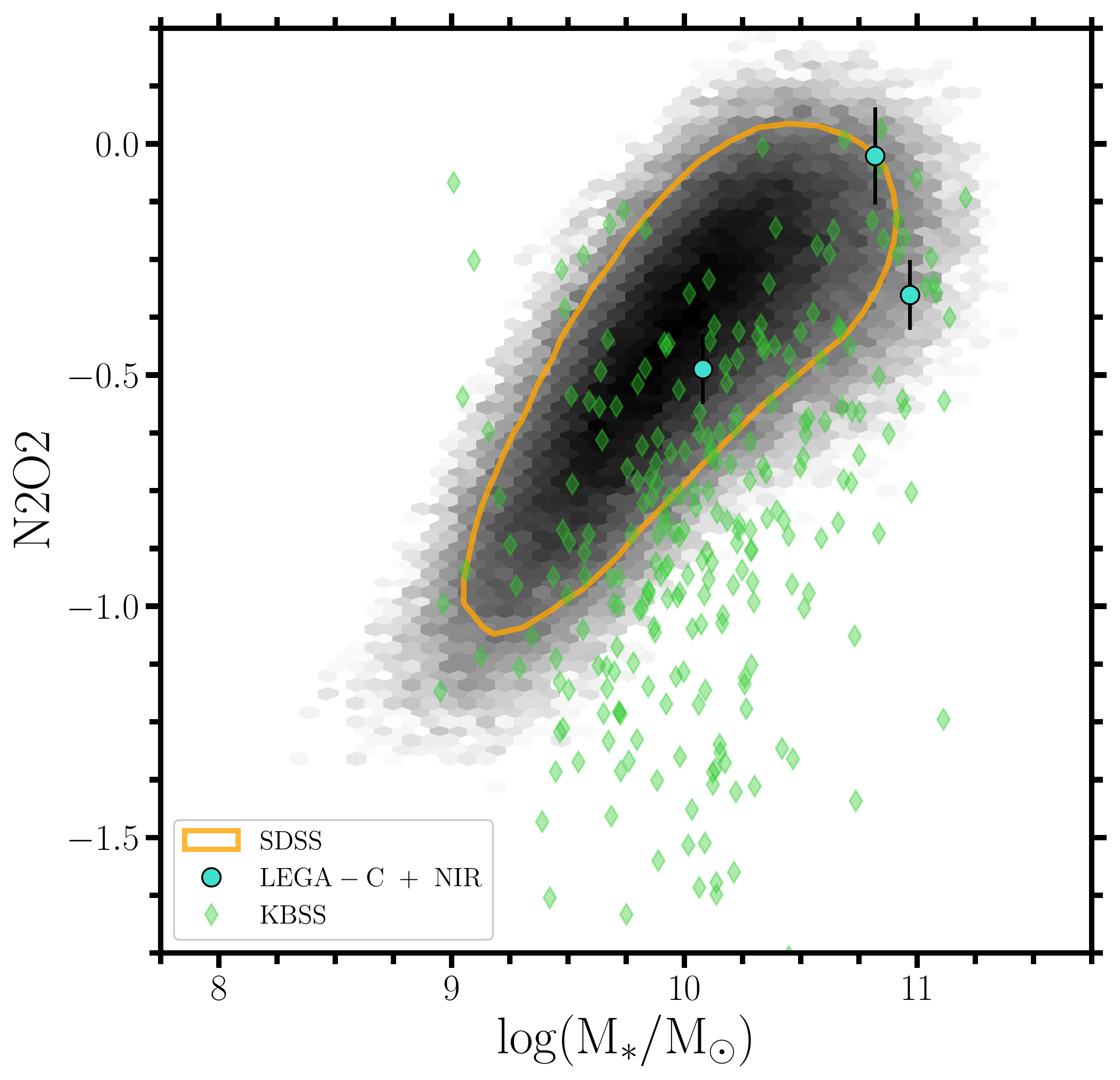}
    \caption{The $\mathrm{M}_{\ast}$-N2O2 diagram. The three $0.6 \lesssim z \lesssim 1.0$ LEGA-C galaxies from Section~\ref{Subsample} with observed-NIR spectra that allow for N2O2 to be determined are shown with the cyan circles. The low-redshift comparison sample of $z \sim 0$ SDSS galaxies from Section~\ref{SDSS} is shown in grayscale with an orange contour, where the contour encloses roughly 90\% of the low-redshift sample. The high-redshift comparison sample of $z \sim 2$ KBSS galaxies from Section~\ref{KBSS} is shown with the green diamonds. We see that the $0.6 \lesssim z \lesssim 1.0$ LEGA-C galaxies have N2O2 intermediate between the $z \sim 0$ SDSS galaxies than the $z \sim 2$ KBSS galaxies.}
    \label{fig:N2O2_mass}
\end{figure}

In Figure~\ref{fig:N2O2_mass}, we present the $\mathrm{M}_{\ast}$-N2O2 diagram. The small subsample of $0.6 \lesssim z \lesssim 1.0$ LEGA-C galaxies from Section~\ref{Subsample} with measurements from observed-NIR spectra is shown with the cyan circles. The low-redshift comparison sample of $z \sim 0$ SDSS galaxies from Section~\ref{SDSS} is shown in grayscale with an orange contour, where the contour encloses roughly 90\% of the low-redshift sample. The high-redshift comparison sample of $z \sim 2$ KBSS galaxies from Section~\ref{KBSS} is shown with the green diamonds. Like the $\mathrm{M}_{\ast}$-O32 and $\mathrm{M}_{\ast}$-O/H diagrams, we exclude galaxies that were identified as AGN from all samples. \par

There are only three LEGA-C galaxies with measurements of N2O2 that are not AGN. Still, we see that these galaxies appear to have line ratios (and, thus, N/O) intermediate between the loci of $z \sim 0$ SDSS and $z \sim 2$ KBSS galaxies. This is consistent with the nebular enrichment results from Section~\ref{SectionThreeThree} and Figure~\ref{fig:Z_R23}. However, the small number of LEGA-C galaxies shown here emphasizes the need for larger samples of intermediate-$z$ galaxies with full rest-optical spectra to understand the role of metallicity. \par

% N2O2-mass

\vspace{8mm}
\subsubsection{The N2-BPT Diagram}
\label{SectionFourFourTwo}

The N2-BPT diagram, which shows O3 versus N2, is often used to determine the dominant ionizing source in galaxies and to separate star-forming galaxies from AGN. As we noted in Section~\ref{SectionThreeOne}, differences in the N2-BPT diagram can be attributed to a variety of astrophysical differences, including differences in (1) ionizing radiation fields, (2) gas-phase metallicity, (3) ISM densities, and/or (4) N/O. Some of these differences we have been able to investigate using other methods. Because the N2-BPT diagram includes line ratios with lines that are close in wavelength, nebular extinction corrections are not necessary, which provides a practical advantage compared to those methods for samples where it can be difficult to accurately determine reddening due to dust. \par

\begin{figure}
    \centering
	\includegraphics[width=0.95\linewidth]{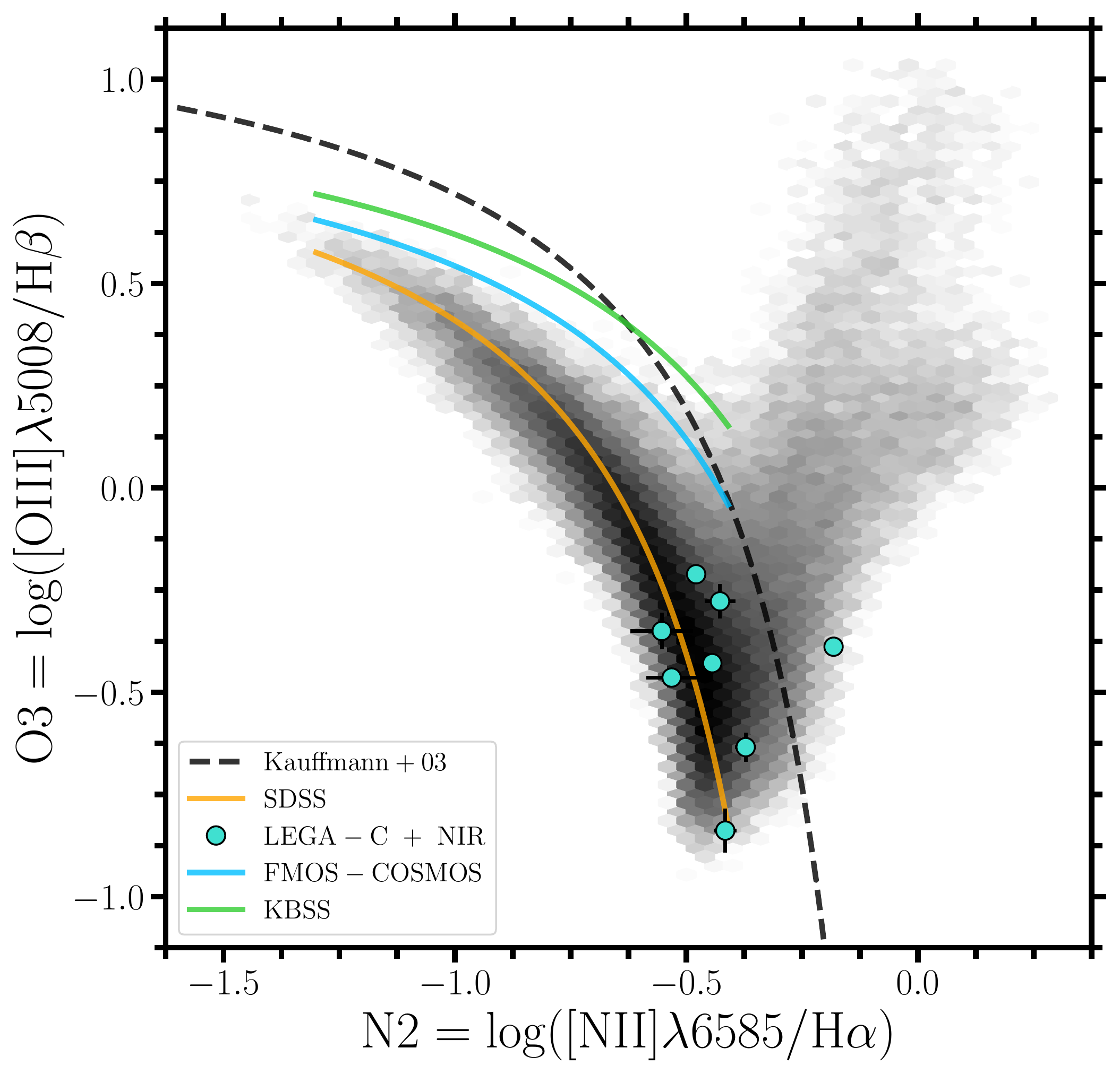}
    \caption{The N2-BPT diagram. The subsample of $0.6 \lesssim z \lesssim 1.0$ LEGA-C galaxies from Section~\ref{Subsample} with observed-NIR spectra is shown with the cyan circles. The low-redshift comparison sample of $z \sim 0$ SDSS galaxies from Section~\ref{SDSS} is shown in grayscale with an orange curve. The intermediate-redshift comparison sample of $z \sim 1.6$ FMOS-COSMOS galaxies from Section~\ref{FMOS} is shown with the blue curve. The high-redshift comparison sample of $z \sim 2$ KBSS galaxies from Section~\ref{KBSS} is shown with the green curve. We see again that the $0.6 \lesssim z \lesssim 1.0$ LEGA-C galaxies appear more similar to the $z \sim 0$ SDSS galaxies than $z \sim 1.6$ FMOS-COSMOS galaxies and the $z \sim 2$ KBSS galaxies.}
    \label{fig:BPT_N2}
\end{figure}

In Figure~\ref{fig:BPT_N2} we present the N2-BPT diagram. The subsample of $0.6 \lesssim z \lesssim 1.0$ LEGA-C galaxies from Section~\ref{Subsample} with measurements from observed-NIR spectra is shown with the cyan circles. The low-redshift comparison sample of $z \sim 0$ SDSS galaxies from Section~\ref{SDSS} is shown in grayscale with an orange curve. The intermediate-redshift comparison sample of $z \sim 1.6$ FMOS-COSMOS galaxies from Section~\ref{FMOS} is shown with the blue curve. The high-redshift comparison sample of $z \sim 2$ KBSS galaxies from Section~\ref{KBSS} is shown with the green curve. \par

The nine $0.6 \lesssim z \lesssim 1.0$ LEGA-C galaxies here have relatively high $\mathrm{M}_{\ast}$ and appear very similar to the $z \sim 0$ SDSS galaxies, as we saw at $10.50 < \mathrm{log}(\mathrm{M_{\ast}/M_{\odot}}) < 11.00$ in the MEx in Section~\ref{SectionThree}. At lower $\mathrm{M}_{\ast}$, $z \sim 0.8$ DEEP2 galaxies with full rest-optical coverage appear less similar to the $z \sim 0$ SDSS galaxies, with minor offsets toward higher values of O3 and/or N2 \citep[][]{Hirtenstein:2021}, as we saw at $10.00 < \mathrm{log}(\mathrm{M_{\ast}/M_{\odot}}) < 10.50$ in the MEx in Section~\ref{SectionThree}. Together, these provide a consistency check between analyses of intermediate-$z$ galaxies with full rest-optical spectra and those with partial rest-optical spectra. \par

It is interesting to consider the combination of physical conditions that would result in $z \sim 1$ galaxies appearing almost identical to $z \sim 0$ galaxies in the MEx, $\mathrm{M}_{\ast}$-O/H, $\mathrm{M}_{\ast}$-N2O2, and N2-BPT diagrams, while also having evidence of somewhat higher ionization as we saw in the $\mathrm{M}_{\ast}$-O32 diagram. In other words, how should we interpret changes in O32 that are not observed in O3, R23, and in the combination of N2 and O3? This likely requires differences in ionization that are not accompanied by differences in metallicity, which could suggest that the relationship between those properties is changing with redshift. Confirming or rejecting this hypothesis will require much larger samples of intermediate-$z$ galaxies across all $\mathrm{M}_{\ast}$, including galaxies at lower $\mathrm{M}_{\ast}$ than explored by LEGA-C, spanning a range in both metallicity and ionization. Assembling such samples is one of the science goals of the upcoming galaxy surveys that will be conducted using both Subaru/PFS and VLT/MOONS. \par

% BPT-N2

%%%%%
% \vspace{8mm}
\section{Summary and Conclusions}
\label{SectionFive}

We have presented a detailed study of the partial rest-optical ($\lambda_{\mathrm{obs}} \approx 3600-5600$\AA) spectra of $N = 767$ star-forming galaxies at $0.6 < z < 1.0$ obtained with VLT/VIMOS as part of the Large Early Galaxy Astrophysics Census (LEGA-C). The large number and high quality of these spectra allowed us to systematically study the emission line properties of this intermediate-$z$ sample of galaxies in the context of other large samples of galaxies in the local Universe with the Sloan Digital Sky Survey (SDSS) and at higher redshifts with the Fiber Multi-Object Spectrograph (FMOS)-COSMOS Survey and the Keck Baryonic Structure Survey (KBSS). Our findings can be summarized as follows. \par

\begin{enumerate}
    \item Using the stellar mass-excitation diagram (MEx), we found that $0.6 < z < 1.0$ LEGA-C galaxies already appear much more similar to $z \sim 0$ SDSS galaxies than $z \sim 1.6$ FMOS-COSMOS or $z \sim 2$ KBSS galaxies (see Figure~\ref{fig:MEX_O3}). There is some evidence that LEGA-C galaxies at low $\mathrm{M}_{\ast}$ ($10.00 < \mathrm{log}(\mathrm{M_{\ast}/M_{\odot}}) < 10.50$) exhibit slight differences with respect to a mass-matched sample of $z \sim 0$ SDSS galaxies (see Figure~\ref{fig:MEX_O3_bins_CDF}), but this occurs near the mass-completeness limit of LEGA-C. However, this mirrors results from a study of $z \sim 1.6$ FMOS-COSMOS galaxies that found higher $\mathrm{M}_{\ast}$ galaxies had more similarities with $z \sim 0$ galaxies than lower $\mathrm{M}_{\ast}$ galaxies at the same redshifts \citep{Kashino:2019}.
    \item To quantify these differences in the MEx diagram, we examined statistical measurements and tests in bins of stellar mass (see Tables~\ref{tab:O3_statistics} and \ref{tab:O3_tests}). When compared to $z \sim 1.6$ FMOS-COSMOS and $z \sim 2$ KBSS galaxies, $0.6 < z < 1.0$ LEGA-C galaxies are significantly offset toward lower values of O3 at all $\mathrm{M}_{\ast}$. When compared to a mass-matched sample of $z \sim 0$ SDSS galaxies, LEGA-C galaxies have significantly different distributions of O3 at lower $\mathrm{M}_{\ast}$ ($10.00 < \mathrm{log}(\mathrm{M_{\ast}/M_{\odot}}) < 10.50$) but not at higher $\mathrm{M}_{\ast}$ ($10.50 < \mathrm{log}(\mathrm{M_{\ast}/M_{\odot}}) < 11.00$). This suggests that the ISM conditions of most galaxies at a lookback time of $6-8$~Gyr are already very similar to present-day galaxies with the same stellar masses, but that there must have been significant changes in the preceding $3-4$~Gyr.
    \item Using O32 and R23 as probes for nebular ionization and enrichment, respectively, we found marginally higher ionization parameter ($\simeq 0.1 - 0.2$~dex higher O32) and similar gas-phase oxygen abundance ($\lesssim 0.1$~dex difference in R23) in LEGA-C galaxies relative to SDSS galaxies at fixed $\mathrm{M}_{\ast}$ (see Figures~\ref{fig:O32_mass} and \ref{fig:Z_R23}).
    \item Looking at trends in emission line luminosities and $K$-band magnitudes within the MEx, we found that selection effects do not have a significant impact on the lower envelope of the $0.6 < z < 1.0$ LEGA-C locus in the MEx, where we see slight differences with respect to a mass-matched sample of $z \sim 0$ SDSS galaxies at lower $\mathrm{M}_{\ast}$ (see Figure~\ref{fig:MEX_O3_luminosity}). Thus, the differences at these $\mathrm{M}_{\ast}$ are likely caused primarily by physical effects. We attempt to assess the impact of AGN at these $\mathrm{M}_{\ast}$ by comparing the locations of galaxies with demarcation lines on the MEx and looking at X-ray luminosities. We suspect that contributions from active galactic nuclei are minor at lower $\mathrm{M}_{\ast}$, with an increasing AGN fraction at higher $\mathrm{M}_{\ast}$ determined using the MEx and X-ray observations.
    \item For a subsample of galaxies that have full rest-optical spectra from new near-infrared observations, we presented comparisons with the $\mathrm{M}_{\ast}$-N2O2 diagram (Figure~\ref{fig:N2O2_mass}) and the N2-BPT diagram (Figure~\ref{fig:BPT_N2}). Although the sample size is small, we found that the $0.6 < z < 1.0$ LEGA-C galaxies also appear more similar to $z \sim 0$ SDSS galaxies than $z \sim 1.6$ FMOS-COSMOS or $z \sim 2$ KBSS galaxies in these parameter spaces.
\end{enumerate}

Using the partial rest-optical nebular spectra of intermediate-$z$ star-forming galaxies, we were able to study the nebular properties of these galaxies in the context of other larger samples of galaxies in the local Universe and at higher redshifts. While our results suggest that many galaxies at a lookback time of $6-8$~Gyr are already very similar to present-day galaxies, the differences between our LEGA-C sample and the $z \gtrsim 1.6$ galaxies from FMOS-COSMOS and KBSS imply that there must have been significant changes in the galaxy population in the $3-4$~Gyr leading up to $z \sim 1$. Clearly, this period in the Universe's history is key to understanding how galaxies transition from highly star-forming to relatively quiescent. \par

In coming years, massively-multiplexed fiber-fed spectrographs will be deployed on large telescopes with Subaru's Prime Focus Spectrograph \citep[Subaru/PFS;][]{Takada:2014} and the Very Large Telescope's Multi-Object Optical and NIR Spectrograph \citep[VLT/MOONS;][]{Taylor:2018}. These spectrographs have wavelength coverage that includes both optical and NIR band passes ($\lambda_{\mathrm{obs}} \approx 3800-12000\,$\AA\ for Subaru/PFS; $\lambda_{\mathrm{obs}} \approx 8000-18000\,$\AA\ for VLT/MOONS), allowing for complete rest-optical coverage for intermediate-$z$ galaxies ($0.5 < z < 1.6$). The surveys planned for these instruments will yield the largest galaxy samples at these redshifts and help determine how and when galaxies were changing during ``cosmic afternoon.'' \par

\vspace{4mm}
% \begin{acknowledgments}

We would like to thank N.~Morrell for acquiring some of the data used here and C.~Steidel, B.~Andrews, and Z.~Lewis for useful discussions that ultimately improved this paper. This work was supported in part by a NASA Keck PI Data Award, administered by the NASA Exoplanet Science Institute (NExScI). The authors particularly wish to recognize and acknowledge the very significant cultural role and reverence that the summit of Maunakea has always had within the indigenous Hawaiian community. We are most fortunate to have the opportunity to conduct observations from this mountain, as we are from the mountain observatories in northern Chile. \par

Funding for SDSS-III has been provided by the Alfred P. Sloan Foundation, the Participating Institutions, the National Science Foundation, and the U.S. Department of Energy Office of Science. The SDSS-III web site is http://www.sdss3.org/. \par

SDSS-III is managed by the Astrophysical Research Consortium for the Participating Institutions of the SDSS-III Collaboration including the University of Arizona, the Brazilian Participation Group, Brookhaven National Laboratory, Carnegie Mellon University, University of Florida, the French Participation Group, the German Participation Group, Harvard University, the Instituto de Astrofisica de Canarias, the Michigan State/Notre Dame/JINA Participation Group, Johns Hopkins University, Lawrence Berkeley National Laboratory, Max Planck Institute for Astrophysics, Max Planck Institute for Extraterrestrial Physics, New Mexico State University, New York University, Ohio State University, Pennsylvania State University, University of Portsmouth, Princeton University, the Spanish Participation Group, University of Tokyo, University of Utah, Vanderbilt University, University of Virginia, University of Washington, and Yale University. \par

% \end{acknowledgments}

\facilities{Keck I (MOSFIRE), Magellan Baade (FIRE), Sloan (BOSS), VLT Melipal (VIMOS)}

\software{\texttt{AstroPy} \citep[][]{Astropy:2013, Astropy:2020}, \texttt{LMFIT} \citep[][]{LMFIT}, \texttt{Matplotlib} \citep[][]{Matplotlib:2007}, \texttt{MPFIT} \citep[][]{MPFIT}, \texttt{NumPy} \citep[][]{NumPy:2011, NumPy:2020}, \texttt{SciPy} \citep[][]{SciPy:2020}}

\clearpage
\bibliographystyle{aasjournal}
\bibliography{main}{}

%% End of the document.
\end{document}